\documentclass{scrartcl}

\usepackage{amsmath}
\usepackage{xargs}
\usepackage{ifthen}
\usepackage{dsfont}
\usepackage[disable]{todonotes}
\usepackage{amssymb}
\usepackage{bm} 
\usepackage{stmaryrd}
\usepackage{pgfplots}
\usepackage{subfig}
\usepackage{listings}
\usepackage{algorithm}
\usepackage{algpseudocode}
\usepackage{hyperref}
\usepackage{multirow}
\usepackage[sort]{cite}
\usepackage{slashbox}
\usepackage{longtable}
\usepackage{eurosym}

\newcommand{\R}{\mathds{R}}
\newcommand{\Z}{\mathds{Z}}

\newcommand{\F}{\mathfrak{F}}

\newcommand{\eps}{\varepsilon}
\newcommand{\ND}{\mathcal{N}}
\newcommand{\set}[1]{\left\{#1\right\}}
\newcommand{\abs}[1]{\left|#1\right|}

\newcommandx{\E}[2][2=\empty]{\ifthenelse{\equal{#2}{\empty}}{\mathrm{E}}{\mathrm{E}_{#2}}\!\left(#1\right)}

\newcommand{\psfragsize}{\small}

\newcommandx{\psf}[3][3=c]{\psfrag{#1}[#3][]{$\psfragsize{#2}$}}
\newcommandx{\psft}[3][3=c]{\psfrag{#1}[#3][]{\psfragsize{#2}}}

\renewcommand{\vec}{\bm}

\newtheorem{thm}{Theorem}[section]{\bfseries}{\rmfamily}
{\bfseries}{\rmfamily}
{\bfseries}{\rmfamily}
{\bfseries}{\rmfamily}
{\bfseries}{\rmfamily}
\newtheorem{exa}[thm]{Example}{\bfseries}{\rmfamily}
\newtheorem{rem}[thm]{Remark}{\itshape}{\rmfamily}

{\bfseries}{\rmfamily}
{\bfseries}{\rmfamily}

\newenvironmentx{proof}[1][1=\empty]{\ifthenelse{\equal{#1}{\empty}}{\emph{Proof.}~}{\emph{Proof (#1).~}}}{\hfill$\square$\\[0.1\baselineskip]}

\definecolor{mygreen}{RGB}{28,172,0} 
\definecolor{mylilas}{RGB}{170,55,241}
\title{On Game-Theoretic Risk Management (Part Three)}
\subtitle{Modeling and Applications}
\author{Stefan Rass
\thanks{Universit\"{a}t Klagenfurt, Institute of Applied Informatics,
System Security Group, Universit\"{a}tsstrasse 65-67, 9020 Klagenfurt, Austria.
This work has been done in the course of consultancy for the EU Project
HyRiM (Hybrid Risk Management for Utility Networks; see https://hyrim.net),
led by the \emph{Austrian
Institute of Technology} (AIT; www.ait.ac.at). See the acknowledgement section.}\\
\texttt{stefan.rass@aau.at}}

\begin{document}

\maketitle

\begin{abstract}
\begin{center}
\textbf{Abstract}
\end{center}
The game-theoretic risk management framework put forth in the precursor
reports ``Towards a Theory of Games with Payoffs that are
Probability-Distributions''
(\href{http://arxiv.org/abs/1506.07368}{arXiv:1506.07368 [q-fin.EC]}) and
``Algorithms to Compute Nash-Equilibria in Games with Distributions as
Payoffs'' (\href{https://arxiv.org/abs/1511.08591}{arXiv:1511.08591v1
[q-fin.EC]}) is herein concluded by discussing how to integrate the
previously developed theory into risk management processes. To this end, we
discuss how loss models (primarily but not exclusively non-parametric) can
be constructed from data. Furthermore, hints are given on how a meaningful
game theoretic model can be set up, and how it can be used in various
stages of the ISO 27000 risk management process. Examples related to
advanced persistent threats and social engineering are given. We conclude
by a discussion on the meaning and practical use of (mixed) Nash equilibria
equilibria for risk management.
\end{abstract}

\pagebreak

\tableofcontents \newpage
\section{Introduction}

With algorithmic matters of game-theoretic risk management being covered in
\cite{Rass2015b},~
it remains to discuss a few (among more existing) possibilities of how game
models can be used in risk management.

First, observe that matrix games directly cover plain optimization as the
special case of an either $n\times 1$ or $1\times m$ game. Both have
applications in risk management, such as helping with the following common
subtasks:
\begin{itemize}
  \item If the current security configuration is to be assessed against a
      number $m$ of (new) threats, we can think of the defender having only
      1 strategy (the current state). The equilibrium in terms of the
      $\preceq$-ordering is then the most severe threat (since the attacker
      maximizes).
  \item Likewise, if several options for mitigating a particular threat are
      available, then the equilibrium being the $\preceq$-minimum
      determined by the defender, is the action leaving the $\preceq$-least
      damage when being mounted.
  \item The general case of $n>1$ and $m>1$ strategies for both, the
      defender and attacker, is discussed in the remainder of this article.
\end{itemize}

\paragraph{Unique Selling Points:}
The method described in the following offers a variety of advantages relative
to the ``conventional'' approach to managing risks, which is usually tied to
intensive discussion (meetings), and also up to difficult consensus finding.
On the contrary, we propose a method that is based on questioning experts
individually, separately, asynchronously and anonymously, which entails the
following features, among well known improvement of the so-obtained data
quality \cite{Perreault.1989}:

\begin{enumerate}
\item Distributed expert interviews that \emph{do not} require to be at a
    certain room at a certain time (no meetings), and thus allow provide
    input in between the normal workflow (asynchronously to the input of
    other experts).
\item Since the questioning is done individually, it can be done
    anonymized. This avoids social or cultural effects that may change a
    person's statement spoken out loud in presence of certain other people
    (superiors, subordinates, etc.)
\item Exploitation of matrix organization: it may well be the case that
    experts are well informed about certain aspects of a problem, but have
    no reliable clue on some other aspects. Polling people by online
    questionnaires in the privacy of their own office allows them to answer
    only those parts of the survey that they can offer input for, while
    leaving them a safe way of refraining from answering other questions.
    Since our method works on probability distributions, the resulting
    dataset from which these are compiled may be richer or sparser,
    depending on how many informed opinions are available. In any case,
    however, asking people face-to-face in a meeting can bring up an
    uninformed guess just to have said something, so that the overall data
    quality is not necessarily as good as in a distributed and individual
    interview.
\item Enlargement of the opinion pool: without the need for a personal (and
    confidential) meeting, people even outside the company can be included
    in the risk assessment. For example, matters of reputation can more
    reliably be assessed by customers (which are typically not invited to
    internal management meetings).
\item The use of distributions also avoids problems of consensus finding or
    opinion pooling (see, e.g.,
    \cite{Jayram04generalizedopinion,Carvalho2013} to mention only two out
    of many publications in this area) towards a representative number. It
    is in that sense preserving all information, since all opinions (all
    available data) goes into the decision process with equal weight.
\end{enumerate}

As a final pro that a game theoretic risk assessment allows, it is even
possible to offer the game solution algorithms as a webservice, where the
modeling and threat assessment can be left to the customer enterprise. For
example, if the client has identified a set of threats and countermeasures,
it will prefer to \emph{not} disclose this information to a subcontractor in
charge of risk management. Since the game model and solution can be
formulated using abstract names only, it is possible to anonymize the data by
letting the threats be only named ``T1'', ``T2'', \ldots, ``Tn'', as well as
the countermeasures be called ``C1'', ``C2'', \ldots, ``Cm''. The real
meaning of threats and countermeasures can then remain private information of
the company, whereas the game model being given in terms of these abstract
identifiers remains solvable by any third party contractor. This allows the
method to be offered as a (web-)service, without running into troubles of
unwanted information disclose (even the number of threats and countermeasures
can be disguised, if a company adds dummy copies of actions to the list to
make it longer than it actually is).

\section{Game-Theory based Risk Management}\label{sec:risk-man-process}


An eloquent and detailed comparison of how game theory fits into and aids the
classical risk management process has been given by \cite{Rajbhandari2011}.
We follow their presentation hereafter, while instantiating and adapting the
specific steps to the concrete setting outlined in the precursor parts to this work \cite{Rass2015,Rass2015b}.
To get started, consider the classical ISO risk management process as
depicted in Figure 1, and compare it to the workflow to be completed when a
game-theoretic model is to be set up; shown in Figure
\ref{fig:game-modeling}.

\begin{figure}
  \centering
  \includegraphics[scale=0.7]{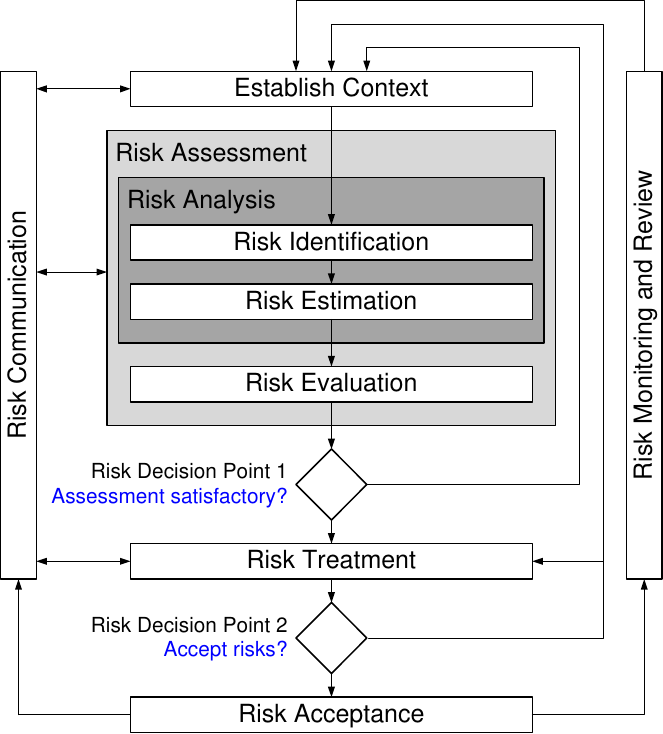}
  \caption{ISO/IEC 27005 Risk Management Process \cite{ISO2011}}\label{fig:iso-risk}
\end{figure}

\begin{figure}
  \centering
  \includegraphics[scale=0.7]{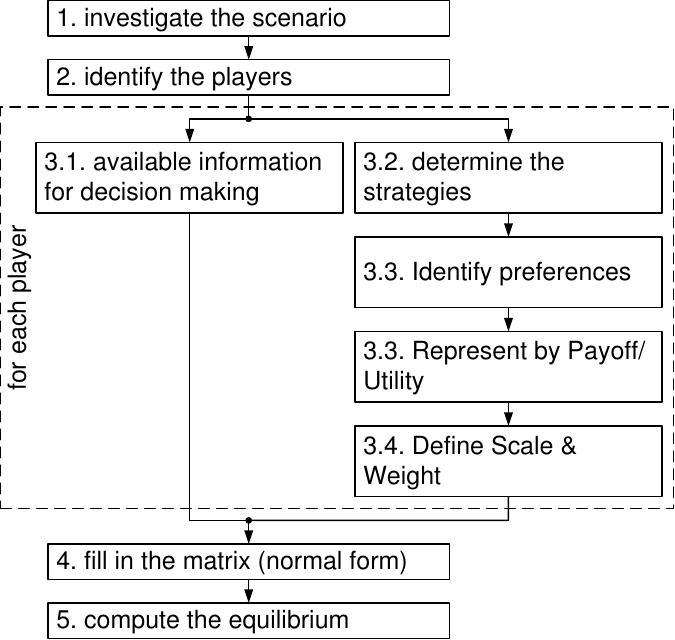}
  \caption{Workflow of setting up a Game Theoretic Model (cf. \cite{Rajbhandari2011})\label{fig:game-modeling}}
\end{figure}

The workflow in Figure \ref{fig:game-modeling} needs some explanation, in
order to establish a mapping to the risk management process. Perhaps the most
important difference between risk management and game theory is the former
being about minimization of losses caused by a not necessarily rational
opponent (nature, but possibly also a hostile party that has explicit
intentions). Contrary to this, game theory in any case assumes a rational
opponent, whose goal is maximizing the own revenue. A conflict/competition
arises if the revenues for both players are negatively correlated, in the
extreme case culminating in the well-known zero-sum competition, meaning ``my
gain is your loss'' (and vice versa). This is the scenario that we also
assume for risk management based on distribution-valued game theory, although
bearing in mind that the incidents that the risk management refers to may not
always follow a hidden rationale or agenda. Nevertheless, the zero-sum
assumption (even though perhaps unrealistic) provides us with a valid
worst-case assessment, and reality can only look better than predicted for
the risk manager. The players engaging in the risk management process may be
diverse and many, depending on the variety of threats to be considered. In
mapping this to a game-theoretic model, we collect all physically existing
opponents into a single adversary acting as ``player 2'', against the risk
manager, which is player 1. As for the adversary, ``player 1'' is here an
abbreviation and means the entirety of people engaged in the practicalities
(``do''-phase of the ISO PDCA-cycle) of the risk management. We will
hereafter call this player the ``defender'', to ease our wording.
Specifically the ``do'' phase is the one where game theory can help, since it
requires:

\begin{itemize}
  \item The selection of controls: this is an action undertaken in the
      final ``risk treatment'' phase in Figure \ref{fig:iso-risk}, and the
      point where game theory is applied.
  \item The implementation of controls: while the selection of controls can
      be based on the knowledge of an equilibrium, the enforcement thereof
      (i.e., playing the equilibrium) corresponds to an optimized
      implementation scheme. We will revisit this issue later in sections
      \ref{sec:tool-support} and
      \ref{sec:optimization-of-the-infrastructure}.
  \item The definition of measures to check their effectiveness: the
      measures per goal are the assurances defined in \cite[Def.4.1]{Rass2015}.
\end{itemize}
So, all these actions can be supported by game theory. To make this precise,
let us look at the steps numbered in Figure \ref{fig:game-modeling}, and to
be completed for each player:

\begin{description}
  \item[3.1:] each player has some a-priori or current knowledge when a
      decision is made. In classical game theory, the next action depends
      only on the current state of the game (in a generalization to
      stochastic or sequential games, a dependency on past game iterations
      is included; we discuss one such example in Section
      \ref{sec:apt-modeling}). In the reality of risk management, external
      need to be considered, and the player's action is hardly dependent on
      the current state of the system only (at least because the system
      state may not even be known precisely at all times). For this reason,
      the
      theory developed in \cite{Rass2015},
      unlike classical game theory, allows uncertainty to be explicitly
      modeled at this stage, and incarnate through the payoffs of an action
      that we describe below.
  \item[3.2:] strategies for a player refer to everything that can be done
      in the current position. In fact, if the game play covers several
      stages until the payoff is received, then the strategy is an exact
      prescription of what is to be done at each step. In that sense, it is
      comparable to a ``recipe'' that the player can follow to accomplish
      the desired goal. Mapping this to a practical setting, the strategy
      may be named ``patch machine X'', whereas its details relate to all
      the steps taken from the current state of machine X until the point
      where the patch has been installed and the machine is put back to
      work. Likewise, a strategy for the opponent (player 2, adversary) may
      be high-level named ``hack machine X'', where the details of this
      strategy may include a sequence of steps such as ``sending a phishing
      mail'' $\to$ ``connect to the malware'' $\to$ \ldots
  \item[3.3:] Identify the interests of a player and define measures of
      ``fulfilment'' of these needs. In our case, this process refers to
      the identification of goals that are object of the risk management.
      Examples include (but are not limited to): economic, reputation,
      people, information, capability, etc. (see \cite{Talbot2011}). The
      degree of achievement in each of these goals must be measured on a
      scale that makes the different goals comparable, and – for technical
      reasons – also arithmetically compatible. The easiest way of assuring
      this is to define a common set of discrete risk categories, with
      individually specific meanings per security goal. This has several
      advantages beyond pure theoretical reasons, as it ``equalizes'' the
      understanding of risk valuations and the taxonomies in which risk and
      outcome of actions is expressed. This is the fundament to the next
      step 3.4.
  \item[3.4:] Preferences of the players are found by asking how the
      players valuate an outcome. For general games, this has to be done
      for each player. In our case (and every zero sum game), it suffices
      to ask only player 1 (the defender) for this valuation, in each
      scenario. By construction, we specifically allow an outcome of a
      specific scenario of strategies for the defender and the attacker to
      be rated in various respects, i.e., in terms of each security goal.
      Having a common vocabulary (set of risk categories) in which the risk
      in each goal is expressed, the underlying theory of multi-criteria
      optimization meaningfully applies and helps optimizing the actions
      towards maximizing the security goal fulfilment (equivalently,
      minimize the residual risk).
  \item[3.5:] the representation of preferences by a utility function is
      --in classical game theory -- done by specifying a function
      $u_i:PS_1\times PS_2\to \R$ for the $i$-th player, so that each
      scenario of defense (action from the list $PS_1$ of available
      options) and attack (action from $PS_2$, i.e., possible exploits) is
      rated in some real-valued score. The construction in \cite{Rass2015}
      deviates from this definition in allowing the outcome to be not crisp
      but a random probability distribution, thus the function $u_i$ takes
      the form $u_i:PS_1\times PS_2\to \F$, where $\F$ is the set of all
      probability distributions (more precisely their density functions)
      that satisfy the regularity conditions (lower bounded by 1, absolute
      continuity w.r.t. Lebesgue or counting measure; see \cite{Rass2015}).
      The crux of this modification is that:
    \begin{itemize}
      \item Letting $u_i$ be valued in terms of probability distributions
          offers a powerful model to capture uncertainty (cf. our remarks
          in step 3.1 above).
      \item The specific definition of $u_i$ can be made based on
          empirical statistics; that is, we can simply collect many
          domain expert opinions on a specific scenario from $(d,a)\in
          PS_1\times PS_2$, and define the value $u_1(d,a)$ as the
          empirical histogram of this expert survey (we expand this
          approach below in section \ref{sec:expert-surveys}). This has a
          neat effect, since it:
          \begin{itemize}
            \item Preserves all information provided by the experts
            \item Avoids a consensus problem that would normally arise
                from the need to agree on a single representative
                opinion about the risk. In practice, people may be
                unsure and disagreeing to the opinion of others, so
                that conflicts and aggregation of opinions may be
                required. In letting the risk valuation be an entire
                histogram, each input goes into the assessment with the
                same weight, so that no domain expert is ``overruled''
                or has less influence than any other.
          \end{itemize}
          If the system response dynamics is known, on the other hand
          side, then there may not even be a need to poll experts, and a
          simulation of the outcomes under the specific scenario $(d,a)$
          could be imaginable. Percolation theory \cite{Koenig2016a}
          offers one way to do this (among other possibilities).
    \end{itemize}
\end{description}

In assuming a zero-sum competition and a two-player game, the risk management
process outlined here is \textbf{asset-centric}. This is an important core
philosophy of the entire process, since the overall goal is not about
preventing all possible threats, but about making an attack non-economic for
the attacker. This is the main point of applying optimization (i.e., game
theory) here, since we seek to minimize our own losses (measured in terms of
the values of our assets), against whatever an attacker may do.

\begin{table}[p]
\scriptsize
  \centering
\begin{tabular}{|p{0.1\textwidth}|p{0.4\textwidth}|p{0.4\textwidth}|}
\hline
\multicolumn{2}{|c|}{ISO/IEC Process/Terminology} & Game theoretic step/terminology\tabularnewline
\hline
\hline
Context establishment & Setting the basic criteria Defining the scope and boundaries Organization
of an information security risk management (ISRM)  & Scenario investigation (scope definition and asset identification)
Player identification (mostly assigning the role of the defender). \tabularnewline
\hline
\multirow{4}{*}{\parbox{0.1\textwidth}{Risk identification}} & Identification of assets & Included in the scenario investigation\tabularnewline
\cline{2-3}
 & Identification of existing controls & Identify implemented controls, i.e. ``do nothing''
option for the defender. \tabularnewline
\cline{2-3}
 & Identification of vulnerabilities & Options that can be exploited by threats. Included while determining
the strategies of the attacker (player 2). We denote this list as
$PS_{2}$.\tabularnewline
\cline{2-3}
 & Identification of consequences & Identify how players value multiple aspects of outcomes. Identify
preferences (i.e., priorities among different security goals)\tabularnewline
\hline
\multirow{3}{*}{\parbox{0.1\textwidth}{Risk estimation}} & Assessment of consequences & Define a common scale and ranking scheme for all relevant outcomes.\tabularnewline
\cline{2-3}
 & Assessment of incident likelihoods & Computed likelihoods for each strategy for both players. \tabularnewline
\cline{2-3}
 & Level of risk estimation (list of risks with value levels assigned) & Expected outcome for each scenario. This expected value (a real number)
is in HyRiM replaced by an entire distribution function (thus avoiding
information loss due to a ``representative''
centrality measure like the average). \tabularnewline
\hline
Risk evaluation & List of risks prioritized & Prioritize the expected outcome for both of the players\tabularnewline
\hline
\multirow{2}{*}{\parbox{0.1\textwidth}{Risk treatment}} & Risk treatment options are risk reduction, retention, avoidance and
transfer & Strategies (control measures) for the defender can be categorized
into ``static'' (changes applied
to the system that have a permanent effect, e.g., installation of
intrusion detection systems) and ``dynamic''
(actions that need to be repeated in order to retain their effect
on security, e.g., security awareness trainings). The entirety of
controls available is denoted as $PS_{1}$.\tabularnewline
\cline{2-3}
 & Residual risks & Expected outcome of the game. In HyRiM, this is the equilibrium value
distribution, from which all statistically meaningful quantities can
be computed. For example, the mean of this outcome distribution would
correspond to the classical quantitative understanding of risk as
the product of likelihood and impact. Since HyRiM admits multiple
goals to be optimized at the same time, the residual risk is returned
per security goal. The theory coined this output artefact ``assurance\textquotedblright ,
and it is one probability distribution for the losses in each relevant
security goal.\tabularnewline
\hline
Risk acceptance & List of accepted risks based on the organization criteria & Strategies of the defender (based on the organization criteria)\tabularnewline
\hline
Risk communication & Continual understanding of the organization\textquoteright s ISRM
process and results & Strategies of the defender\tabularnewline
\hline
Risk monitoring and review & Monitoring and review of risk factors Risk management monitoring,
reviewing and improving  & The process is repeated as the player\textquoteright s options and
their outcome valuation may change\tabularnewline
\hline
\multicolumn{2}{|c|}{not included} & Information gained by the opponent\tabularnewline
\hline
\multicolumn{2}{|c|}{not included} & Beliefs and incentives of the opponent\tabularnewline
\hline
\multicolumn{2}{|c|}{not included} & Optimization of the strategies\tabularnewline
\hline
\end{tabular}
  \caption{Mapping between ISO/IEC 27005 Risk Management and Game Theoretic Approach (cf. \cite{Rajbhandari2011})}\label{tbl:mapping}
\end{table}

The mapping sketched in Table \ref{tbl:mapping} has been adapted from
\cite{Rajbhandari2011}, but needs a bit of tailoring towards defining an
analogous process based on distribution-valued games (and the theory
thereabout). Specifically, the risk assessment phase (Figure
\ref{fig:iso-risk}), comprising risk identification, risk estimation and risk
evaluation, refers to the current state of the system, whose valuation may
trigger further action (implementation of controls) in later phases of the
process (namely the risk treatment).

\begin{figure}
  \centering
  \includegraphics[scale=0.7]{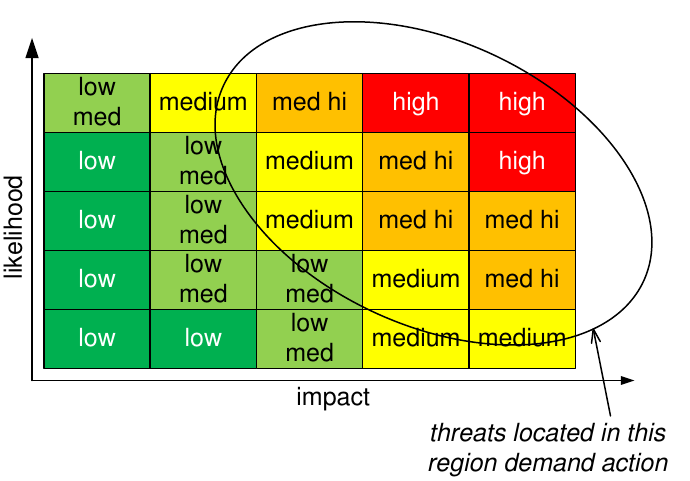}
  \caption{Risk Matrix (example)}\label{fig:risk-matrix}
\end{figure}

In Table \ref{tbl:mapping}, the process of determining the current state is
only briefly mentioned as adding a strategy ``do nothing'' to the game, which
merely models the possibility of the situation at hand being satisfying
already. Here, we will resemble the classical process of risk management to a
wide extent using only the idea of an outcome valuation in terms of
probability distributions, which is made possible by the $\preceq$-ordering
on these objects that was invented in earlier stages of the project. This
extends up to the point where risks are evaluated, which is normally done
using a 2-dimensional risk matrix, typically with colored entries like shown
in Figure \ref{fig:risk-matrix}. The (specification of the) elliptic region
displayed on the upper right corner is the main output of the risk
evaluation, which establishes a priority list of risks and divides the
threats into those that demand action (i.e., which are put into the critical
region based on their impact and likelihood) and those who do not demand
immediate actions (i.e., which lie outside the critical region).

In the original ISO/IEC process, the risk value is computed as the product
$risk=likelihood \times impact$. It is popular to define a categorical scale
for both, the likelihood and the impact (in tabular form explaining what
exactly is meant by a ``medium likelihood'' or a ``medium impact''; section
\ref{sec:loss-categories} gives examples). Multiplying the ranks (i.e.,
category numbers\footnote{it is advisable to avoid using the number zero as a
category number or rank, as it would cancel out any rating in the other
aspect. Indeed, we have adopted such an assumption quite explicitly already
by declaring loss random variables (or categories) to be $\geq 1$ in any
case; see \cite{Rass2015a}.}) then gives the risk score and delivers the
coloring of the risk matrix as exemplified in Figure \ref{fig:risk-matrix}.
Note that loss valuations based on crowd sourcing (expert surveys, detailed
in section \ref{sec:expert-surveys}) in fact embody both, the impact (as is
directly asked for), and the likelihood (as the relative frequency of answers
provided by the set of experts polled) in one object. That is, the classical
two-dimensional matrix depicted in Figure \ref{fig:risk-matrix} would boil
down to a 1-dimensional list of threats, which can plainly be sorted in
$\preceq$-ascending order. Thus, the risk evaluation is even simplified in
our setting here. It may nonetheless be advisable to resemble the classical
process of risk management a little closer by asking for both, the impact and
the likelihood, and construct empirical distributions for both. This may be
useful in letting the expert express his beliefs more detailed than just
asking for a possibility (and not for a probability too). We revisit this
issue later in section \ref{sec:risk-priorization}, after having established
the basics of this process first.

In generalizing this classical approach to risk management to a more
sophisticated technique based on games and distributions as risk representing
objects, our main target in the next section will be \emph{modeling the
losses} as probability distributions, and how to prepare the game theoretic
models for the risk treatment phase, when the games are to be solved and
``played'' in practice. We will exemplify the modeling based on two examples,
which are advanced persistent threats (APTs), covered in section
\ref{sec:apt-modeling}, and social engineering, discussed in section
\ref{sec:human-element}. Once the cycle in Figure \ref{fig:iso-risk} has been
completed for once, any repetition entailing another round of risk estimation
and risk evaluation can be supported by exactly the same kind of
game-theoretic models that were used in the previous risk treatment phase (in
fact, the equilibrium strategy enforced as result of the last risk treatment
phase is the ``do nothing'' strategy in the next risk assessment; cf. Table
\ref{tbl:mapping}). Examples of potentially suitable models are described in
the sections to follow. Section \ref{sec:optimization-of-the-infrastructure}
discusses how to use the game analysis results for effective improvements and
control selection.

\section{Modelling Losses}
The proposed method of risk modeling is here based on non-parametric loss
models, for their conservation of information and absence of perhaps
difficult to verify assumptions. Nonetheless, actuarial science knows
important applications of parametric loss models, which we briefly discuss in
the next section.
\subsection{Parametric Loss Models}
The weight of a distribution's tails is what makes it (in)appropriate for
risk management, where distributions with heavy, fat or long tails are common
choices. In the continuous case, extreme value distributions (Gumbel, Frechet
or Weibull) are suitable choices, as well as stable distributions. The
latter, despite not having analytically expressible densities except for
special cases, can nevertheless be ordered upon using sequences of
truncations~
or moment sequences (if they exist) \cite{Rass2015b}.

In the discrete case, the $(a,b,0)$ or $(a,b,1)$ class of distributions may
be considered, with a density in the former class being defined recursively
as $\Pr(X=k)=\Pr(X=k-1)\cdot(a+\frac b k)$ for $k=1,2,3,\ldots$. The
$(a,b,1)$ class is -- roughly speaking -- the truncated version of an
$(a,b,0)$-density excluding the possibility of the event $X=0$. It can be
shown that the former class includes exactly three families of distributions,
which are Poisson, Binomial and the Negative Binomial distribution.

A general issue with parametric losses is their representation of an
arbitrary amount of information by a fixed number of parameters. This
inevitably incurs a loss of information, and calls for partly sophisticated
methods of parameter fitting or similar. On the contrary, nonparametric
losses like (the previously proposed) kernel densities come with the appeal
of preserving all information upon which they are constructed, as well as
offering the flexibility of allowing for adjustments to model uncertainty in
the expert's answers more explicitly. An example of this will be sketched in
section \ref{sec:bandwidth-certainty}.

Nonetheless, if a parametric model should be used, then one should bear in
mind that risk is intrinsically a latent variable, and the best we can do is
relate it to some observable variables. Unfortunately, the usual assumptions
on the latent variable being dependent on the hidden variable, so that an
inference towards risk is possible, may not directly apply to a risk
assessment process. Risk is a property of some general entity, action,
intention, or similar. As such, subjective assessments of it are made by
humans whose risk perception may be correlated (to a degree that depends on
the skill level of the person in the respective regard) but not directly
influenced by the true underlying objective value of the risk variable.
Still, algorithms like expectation-maximization \cite{McLachlan.2008} could
(and should) be assessed for the extent to which they can deliver a useful
risk model. Indeed, the result of an EM-algorithm being a distribution model
over latent variables together with a point estimate on its parameters is a
perfectly suitable input for the decision models put forth in the precursor
parts of this report.

\subsection{Nonparametric Loss Models}
Parametric models come at the cost of information loss due to representing
the entire data by a (preferably small) set of parameters to the
distribution. This is the price for analytic elegance and the ease of many
matters related to working with such models practically.

In light of the scarceness and inconsistencies in the data to be expected in
a risk management application, we may thus look at nonparametric loss models
as a potentially interesting alternative. The method of choice here will
follow the proposals made in \cite{Rass2015b}.
Specifically, we will start from an empirical distribution compiled from
expert interview data, and use a kernel density estimator for our loss model.
Moreover, we will focus on discrete risk assessment scales, which lead to
discrete (in fact categorical) empirical distributions. For a kernel density
estimator in this context, it is somewhat surprising that the literature is
unexpectedly thin on discrete kernel density estimates. Besides only a few
proposals found in \cite{Rajagopalan1995,Li2003,Chesson2005}, the problem of
``good'' kernel density estimation seems to be as much of an art (on top of
science) as it is in the continuous case. Hence, we will continue defining
our own discrete kernel proposal based on Gaussian densities, but postpone
the details until later, when we have specified the process of data
collection, which obviously goes first.

\subsection{Collecting Data from Experts}\label{sec:expert-surveys}
As for every empirical study, the first step is fixing the details of the
scenario about which experts are to be interviewed. In our context, this may
entail the identification of a particular threat and related countermeasures.
The high level description of a threat, say ``unauthorized access to server
$X$'', must here be followed by a sequence of possible detailed suggestions
on how the respective attack could be mounted. For the aforementioned
example, possibilities include the exploitation of (known) vulnerabilities
(up to zero-day exploits) in the server itself, social engineering, theft of
access credentials, or similar. Normally, each of these possibilities is
matched with a respective countermeasure. Risk management standards like ISO
27000 \cite{ISO2016,ISO2009} or related ones \cite{BSI2008b} provide an
indispensable source of threats and countermeasures, which can be used in
this step.

Abstractly, let us think of this process having brought up a list $PS_1$ of
countermeasures, opposing a list $PS_2$ of possible threat scenarios (the
acronym $PS$ means ``pure strategy'', as a reminder that we are approaching a
game-theoretic model here).

The lists $PS_1$ and $PS_2$ can be assumed to be quite short (though they
must be comprehensive; ideally exhaustive) in practice, and each scenario
$(d,a)\in PS_1\times PS_2$ of ``defense ($d$)-vs-attack ($a$)'' can be put to
its own individual review (not necessarily independent of other scenarios,
but the design of questionnaires and the amount of context specification in
there is a different and nontrivial story of empirical science not subject of
this report).

\begin{rem}[Simultaneous Occurrences of Multiple
Scenarios]\label{rem:multiple-simultaneous-threats}

Note that the modeling so far implicitly prescribes the assessment to be done
relative to a specific single threat. This restriction can be dropped in
presenting the expert a set of threats and explicitly allowing for several of
them to occur at the same time. At first glance, this would exponentially
enlarge the list $PS_2$, this combinatorial explosion can be avoided by
allowing the expert to name multiple possible loss categories with different
likelihoods. That is, the expert may center the thoughts around a specific
threat, but may take further considerations of coincidental other incidents
into account in saying that, for example,
\begin{itemize}
  \item losses of category $x$ are most likely,
  \item while losses of the larger category $y>x$ remain possible if two or
      more incidents occur at roughly the same time.
\end{itemize}
If several such possibilities are uttered, they are most conveniently
described by a distribution supported on (at least) the anticipated loss
categories $x$ and $y$.
\end{rem}

To set up the empirical game theoretic model, let us therefore assume that
one scenario $(d_1,d_2)$ has been specified in detail, say (for illustration)
\begin{tabbing}
  $d_1$: \= perform periodic updates\\
  $d_2$: \> exploit some software vulnerability
\end{tabbing}
It goes without saying that this specific (example) scenario appears (if at
all) in the middle of a real-life APT, as the earlier stages are usually
matters of social engineering to make an initial contact and infection. We
will go into details of this in section \ref{sec:apt-modeling}, and keep our
example abstract here only for the sake of illustration.

\begin{figure}
  \centering
  \includegraphics[scale=0.7]{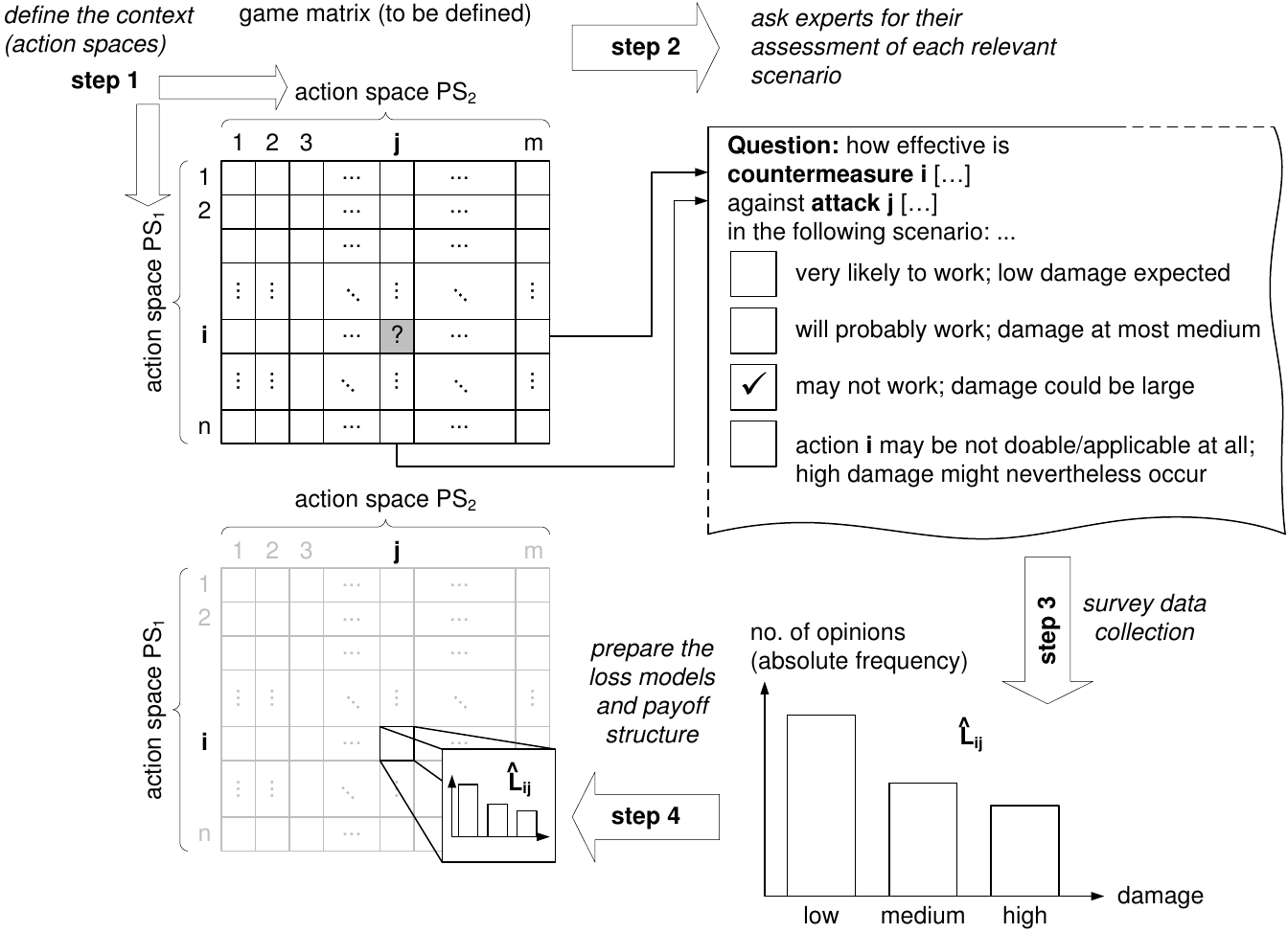}
  \caption{Setting up a Matrix Game with Uncertain Payoffs from Expert Questionnaires}\label{fig:workflow}
\end{figure}

The goal of data collection is getting a payoff matrix composed from loss
distributions that can be used with the game theoretic framework defined in
the preceding reports \cite{Rass2015a,Rass2015b}.
Figure \ref{fig:workflow} displays the workflow to fill in \emph{one} cell in
the payoff matrix, which is basically done along four steps:
\begin{enumerate}
  \item\label{lbl:step1-scenario-description} Selection and specification
      of a scenario $(d_i,d_j)\in PS_1\times PS_2$ as a questionnaire
      presentable to experts (the wording and style of the questions is
      itself a highly nontrivial matter, and should be done w.r.t. the
      subsequent implementation strategy for the optimal defense. We will
      revisit this issue later in section \ref{sec:equilibrium-play}),
  \item\label{lbl:step2-expert-survey} Doing an expert survey on the
      effectiveness of countermeasure $d_i$ against attack $d_j$. It is
      crucial for this survey to clearly define at least the following
      items:
      \begin{itemize}
        \item the context of the risk assessment (that is, the aspects
            that are relevant and those that are irrelevant for the risk
            assessment)
        \item a clear definition of the scale in which the risk is
            quantified (in the style of Table
            \ref{tbl:risk-categories-examples} shown in section
            \ref{sec:loss-categories} or similar).
      \end{itemize}
  \item\label{lbl:step3-loss-distribution-definition} Collecting as much
      expert input as possible to define an empirical loss distribution
      $\hat L_{ij}$ over the categories as specified in the survey.
  \item\label{lbl:step4-loss-distr-preparation} Preparing the loss
      distribution for a subsequent game-theoretic analysis.
\end{enumerate}
The details of step \ref{lbl:step1-scenario-description} are individually
dependent on the application context, i.e., are specific for the system or
infrastructure under investigation. This step is recommended to be done
according to established standard procedures as described by ISO 27000 or its
relatives. Step \ref{lbl:step2-expert-survey} is a matter of empirical
research and questionnaire design. The literature on this is rich, and this
task should be left with experienced staff educated in empirical research and
statistics. On the technical level, collecting the data is relatively simple,
since it is an easy task to set up an online survey, displaying a sequence of
questions asking the expert to give her/his risk rating on each described
scenario. Things may be even made more efficient by showing the expert the
entire matrix, asking to enter a risk assessment in each cell (showing yet
another combination of defense and attack), and to leave all fields blank for
which there is a lack of domain knowledge or no justified opinion can be
expressed.

Step \ref{lbl:step3-loss-distribution-definition} deserves some attention, as
this involves an a-priori agreement on the loss categories to be used in the
survey. We give details on this in section \ref{sec:loss-categories}.

The final preparation of loss distributions in step
\ref{lbl:step4-loss-distr-preparation} is a matter of kernel smoothing, and
described in section \ref{sec:discrete-kernel-smoothing}.

\subsection{Refining the Expert Survey}
Note that the expert survey -- in the form described -- only asks for
possibilities and not for probabilities. That is, the expert is only
questioned to state an expectation of outcome, rather than telling how likely
s/he feels it to be. If the survey is refined to ask for a likelihood in
addition (as is prescribed in the conventional ISO/IEC risk management
standards), then we end up with two probability distributions, one for the
impact, the other for the likelihood. A ``multiplication'' of these two
objects, to resemble the usual formula ``risk $=$ impact $\times$
likelihood'' underlying quantitative risk management, is theoretically
possible (as a multiplication of the hyperreal representatives), but not
practically meaningful. If the two distributions are available, then a
lexicographic comparison of $(impact,likelihood)$ or $(likelihood,impact)$
may be the more reasonable way to go. However, an explicit advantage of the
expert surveys and using distributions as representative objects is them
embodying both, impact and likelihoods within the \emph{same} loss
distribution object, thus the expert interviews are greatly simplified over
the standard risk management process.

\subsection{Outlier Elimination}
It is important to note that any survey data should be cleaned from outliers,
and there has to be a consensus on the treatment of missing values. Either is
up to a variety of statistical methods and the particular method should be
chosen in light of the given application.

The discussion here is only meant to bring this point to the attention of the
reader, and will not go into details of how this could be done.

\subsection{Harmonizing Risk Attitudes}
After having cleaned outliers, it remains to ``harmonize'' different kinds of
answers depending on the individual personalities and risk attitudes. Persons
known to be risk averse will tend to overestimate the risk, while risk
seekers will tend to underestimate. It is beyond the scope of this report to
discuss concrete methods to correct risk estimates based on a respective
classification of the individual, but it is important to bear in mind methods
of statistical classification as a potential toolbox to help in this regard.

\subsection{Using the Bandwidth Parameter to Model Answer
Uncertainty}\label{sec:bandwidth-certainty} In some occasions, it may happen
that an expert is unsure about whether or not certain circumstances enable
certain damages. For example, if the question of whether an incident at one
point in the system can cause damage at another point in the system can be
answered with ``generally no, except for some rare cases'', then the risk
assessment related to the incident will have its modal value at a low
category, but -- due to the possibility of the incident being nonetheless
severe -- extends the distribution up to the full range of loss categories.
The bandwidth parameter of the kernel density estimate can be increased to
let the density put more weight on far away categories, or be chosen smaller
when the certainty about the assessment (rareness of the incident), is
better. In any case, the choice of bandwidth for the kernel density estimates
gains another dimension of importance as being a parameter to
control/describe the expert's certainty in the data.

\subsection{Defining Risk Assessment Categories}\label{sec:loss-categories}
It is crucial (not only for technical reasons) that \emph{all} loss
distributions for \emph{all goals} in a multi-goal security game must be
defined on the \emph{same categories}. This can be justified by technical but
also interpretative reasons:
\begin{description}
  \item[Technical reason:] the procedure to numerically compute multi-goal
      security strategies (MGSS) relies on casting the multi-criteria
      objective function (vector-valued) into a scalar that is a weighted
      sum. This transformation requires ``compatible'' objects to be
      weighted and added (see \cite{Rass2015b}), which calls for the same
      underlying scale in all goals.
  \item[Interpretation:] comparing the severity of damages in two security
      goals is only meaningful if the goals are quantified in the same
      terms. That is, the understanding of, say ``high'' damage has to be
      fixed for all goals, and must not be left to a subjective idea that
      is individual for each expert (otherwise, the outcome of any such
      assessment is useless). This is especially relevant for non-numeric
      goals like the reputation of an enterprise, customer trust, or
      similar.
\end{description}

Especially to address the latter, it appears advisable to underpin the
survey/questionnaire by an a priori fixed definition of categories in which
the risk assessment shall be made. This can be done in tabular form where
each goal is assigned a column, with rows corresponding to the categories,
and cell entries describe the meaning of a risk category specifically for
each goal. Table \ref{tbl:risk-categories-examples} shows an example, whereas
it must be stressed that the concrete content of the table must be
adapted/tailored to the practical situation at hand.

\begin{table}
  \centering
  \scriptsize
  \begin{tabular}{|p{0.15\textwidth}|p{0.23\textwidth}|p{0.23\textwidth}|p{0.23\textwidth}|c|}
\hline
\multirow{2}{*}{\parbox{0.15\textwidth}{risk category (numerical representative)}} & \multicolumn{4}{c|}{loss category}\tabularnewline
\cline{2-5}
 & loss of intellectual property & damage to reputation & harm to customers & \ldots\tabularnewline
\hline
negligible (1) & $<500$ ~\euro & not noticeable & none & \ldots\tabularnewline
\hline
noticeable (2) & between 500~\euro\newline and 10.000 ~\euro & noticeable loss of customers (product substituted) & inconvenience experienced but no physical damage caused & \ldots\tabularnewline
\hline
low (3) & $>10.000$ ~\euro~and 50.000 ~\euro & significant loss of customers & damages to customers' property, but nobody injured & \ldots\tabularnewline \hline medium (4) & $>50.000$~\euro~and
200.000~\euro & noticeable loss of market share (loss of share value at stock
market) & reported and confirmed incidents involving light injuries &
\ldots\tabularnewline \hline high (5) & $>200.000$ ~\euro~\newline and 1
Mio. ~\euro & potential loss of marked (lead) & reported incidents involving
at least one severe injury but with chances of total recovery &
\ldots\tabularnewline \hline critical (6) & $>1$ Mio. ~\euro & chances of
bankrupt & unrecoverable harm caused to at least one customer &
\ldots\tabularnewline \hline
\end{tabular}
  \caption{Potential (Example) Definition of Risk Categories for different Security Goals}\label{tbl:risk-categories-examples}
\end{table}

Such a table should be displayed together with the survey, to equalize the
expert's individual understandings of the risk categories, and to harmonize
the resulting data. The loss categories actually used with the model are
simply the integers $1,2,3,\ldots$, noting that the number 0 is precluded as
a category (in order not to violate the assumptions made in \cite{Rass2015a}). 

\subsection{Using Continuous Scales}
If the loss is measured in continuous terms, then a common categorization
like outlined above must be replaced by a ``meaningful'' common (continuous)
risk range. The exact definition of ``meaningful'' must herein be made
dependent on the context of the problem, so that all security goals (losses)
homogeneously cover the range without being concentrated in disjoint regions.
We illustrate the issue with two examples, one showing how it should be done,
the other illustrating the problem:

\begin{exa}[well chosen loss range]
Consider two loss variables, with are monetary loss due to theft of
intellectual property, and (monetary) investments to protect these assets.
Since security is primarily not about making an attack impossible but only
about making it non-economic, it can be expected that both losses range in
roughly the same numeric region. Formally, let $R_1, R_2$ be the two random
loss variables, and let $\eps>0$ be a small value for which we truncate both
distributions $F_1,F_2$ at their respective $(1-\eps)$-quantiles, denoted as
$q_{(1-\eps)}(F_i)$ for $i=1,2$. Call the so-obtained loss ranges
$I_1=[1,q_{(1-\eps)}(F_1)]$ and $I_2=[1,q_{(1-\eps)}(F_2)]$. If $I_1\approx
I_2$ (i.e., $q_{(1-\eps)}(F_1)\approx q_{(1-\eps)}(F_2)$), then we may take
the convex hull of $I_1\cup I_2$ as the common loss range, and be sure that
the comparativeness of the two loss variables is retained and reasonable.
\end{exa}

\begin{exa}[badly chosen loss range]\label{exa:bad-lossrange}
Let $R_1\sim\ND(5,1), R_2\sim\ND(900,100)$ be two random losses with Gaussian
distributions. As before, if the modeler chooses the range as the (convex
hull) of the union of both (truncated) ranges at some $(1-\eps)$-quantile,
the game may be defined over losses within the common range $[1,1000]$ (for
$\eps=0.027$, the range $\mu\pm 3\sigma$ covers $\approx 99.73\%$ of the
cases for the Gaussian distribution $\ND(\mu,\sigma)$; our cut off here is at
$\mu+5\sigma$, and thus covers more than $99.9\%$). Since $R_2$, however,
assigns most of its mass in the region $[800,1000]$, it will always (at least
up to reasonable numerical precision) be $\preceq$-larger than $R_1$, so the
optimization is pointless.
\end{exa}
To avoid situations like sketched in Example \ref{exa:bad-lossrange}, (at
least) two options are available:
\begin{enumerate}
  \item Rescaling of all loss ranges to a common region. Continuing example
      \ref{exa:bad-lossrange}, this would mean replacing $R_2$ by $\frac 1
      {90} R_2\sim \ND(10,100/(90^2))\approx \ND(10,0.012346)$.
  \item Defining a common continuous scale with the same intended usage as
      the discrete categories. That is, if a continuous loss rating is
      permitted, we may define it within a common interval from $[1,10]$,
      allowing any value to be picked by the modeler, as long as it is
      within the range. This may be the method of choice if the modeling
      interface displays a ``slider'' where a user can drag the gauge at
      any position on a continuous scale to express the (subjective) belief
      about the loss.
\end{enumerate}

\subsection{Preparing the Loss Model}\label{sec:discrete-kernel-smoothing}
In \cite{Rass2015b}, the theoretical 
possibility of convergence issues in the game's analysis (by fictitious play)
was anticipated, occurring when the loss distributions do not share the same
support (this adds to the technical justification stated in the previous
section \ref{sec:loss-categories}). Practically, this is highly likely to
occur due to missing data. If some categories are simply not being used
(either because of an unsuitable definition or because the scale is
fine-grained so that not all levels are being used by the experts), then the
loss model may be empirically correct, but not useful with a subsequent
numeric analysis.

The solution in the continuous case was a kernel smoothing, technically a
convolution with a Gaussian density, to extend all loss distributions until a
common end of their support (previously called the \emph{cutoff point}) is
reached. We will do the same thing in the discrete case, using a kernel
obtained from discretizing the Gaussian distribution. To this end, define
$f(x):=(1/\sqrt{2\pi})\exp(-\frac 1 2 x^2)$ as the density of a standard
normal distribution, and for every $h>0$, define the kernel function
$K_h:\Z\to\R$ as
\[
    K_h(n) := \frac 1 h \int_{n-\frac 1 2}^{n+\frac 1 2}f(t/h)dt,
\]
where $h$ is the bandwidth parameter (as familiar from continuous kernel
density estimates). Obviously, $\sum_{n=-\infty}^\infty K_h(n)=\frac 1
h\int_{-\infty}^\infty f(t/h)dt=1$ and $K\geq 0$, so $K$ defines a discrete
probability mass function on $\Z$. Moreover, it is easy to see that for every
$h>0$, the function $K_h$ resembles a discrete version of a Gaussian density,
so that by letting $h\to 0$, $K_h$ degenerates into a discrete Dirac mass,
\begin{equation}\label{eqn:discrete-kernel-convergence}
  K_h(n)\to \left\{
             \begin{array}{ll}
               1, & \hbox{if }n=0; \\
               0, & \hbox{otherwise.}
             \end{array}
           \right.
\end{equation}
If we call $\hat F_n$ a general empirical distribution function (defined on
any subset of $\Z$) obtained from $n$ data points (answers in the expert
survey). The convolution $\hat F_n\ast K_h$ is another distribution function
supported on all $\Z$, i.e., $(\hat F\ast K_h)(n)>0$ for all $n\in\Z$.
Moreover, in letting $h\to 0$, we have the pointwise convergence $(\hat F\ast
K_h)(n)\to\hat F(n)$ for all $n\in\Z$, so the estimate is asymptotically
correct (note that in contrast to Nadaraja's theorem, we do not even need to
assume a specific speed of decay when letting $h\to 0$). This convergence
easily follows from \eqref{eqn:discrete-kernel-convergence}. The same result
equivalently holds if we replace $f$ by a truncated version $\hat f$ thereof
(being supported on $[1,a]$ for some integer $a>1$), leading by disretization
to the truncated kernel $\hat K_h$. In the following, we thus consider the
smoothed empirical distribution $\hat F_{n,h}:=\hat F_n\ast\hat K_h$, noting
that this also pointwise converges to $\hat F$ and that the support
(strictly) covers the full interval $[1,a]$. However now, we can make a
stronger convergence statement: let $F$ be the (unknown) distribution of the
loss assessment, which is approximated by the expert data\footnote{We
somewhat sloppily assume here that the experts provide us with
``observations'' about the real random loss $X$ having the distribution
function $F$. This assumption is clearly not correct, but still the best that
can be done, given that we cannot simply wait for losses to occur, as this
would probably kill the enterprise much before a decent lot of data about the
true loss distribution could have been obtained. Thus, we have to live with
overly trusted experts providing us with ``objective'' samples of the
unobservable loss variable distributed like $F$.}, which is $F$. We can write

\begin{equation}\label{eqn:loss-distribution-convergence}
  \abs{\hat F_{n,h}-\hat F_n+\hat F_n-F}\leq \abs{\hat F_{n,h}-\hat
F_n}+\abs{\hat F_n-F}
\end{equation}
To the first term, we can apply a uniform bound thanks to the supports of
$\hat F_{n,h}$ and $\hat F_n$ being all finite, which is
\[
    \sup\abs{\hat F_{n,h}-\hat F_n}\to 0\quad\text{ as }\quad h\to 0,
\]
as is an easy consequence of the equivalence of all norms on $\R$. The other
term in \eqref{eqn:loss-distribution-convergence}, we can apply
Glivenko-Cantelli's theorem to conclude the convergence
\[
\sup\abs{\hat F_n-F}\to 0,\quad\text{ as }\quad n\to \infty,
\]
so that $\sup\abs{\hat F_{n,h}-F}\to 0$ in the limit $h\to 0$ and
$n\to\infty$. Note that this argument is only good for \emph{plausibility} of
our approach, but cannot be taken as a proof of correctness, since
practically, it still relies on an infinitude of data, which -- more
crucially -- must be objectively sampled from the real loss variable (having
the distribution $F$).

On the bright side, however, the proposed smoothing enjoys a nice intuitive
justification as accounting for uncertainty in the assessment.

\begin{exa}\label{exa:scoring-scale}
Suppose an expert utters the opinion that, on a scale from 1 to 5, the risk
is 3. Despite that a middle assessment is an implicit statement of
uncertainty already\footnote{This is usually avoided by choosing a risk scale
with an even number of categories to avoid having the median on the scale.},
there statement ``the risk score is 3'' is not equivalent to the statement
that the risk cannot be anything else. In other words, there may be admitted
chances of the damage being higher or lower than the average told by the
expert. To express such uncertainty more detailed, the expert may admit
possible outcomes on the entire range $1,\ldots, 5$, with the initial
assessment being just the most confident outcome.
\end{exa}

\begin{figure}
  \centering
  \subfloat[Original empirical histogram]{\includegraphics[width=0.25\textwidth]{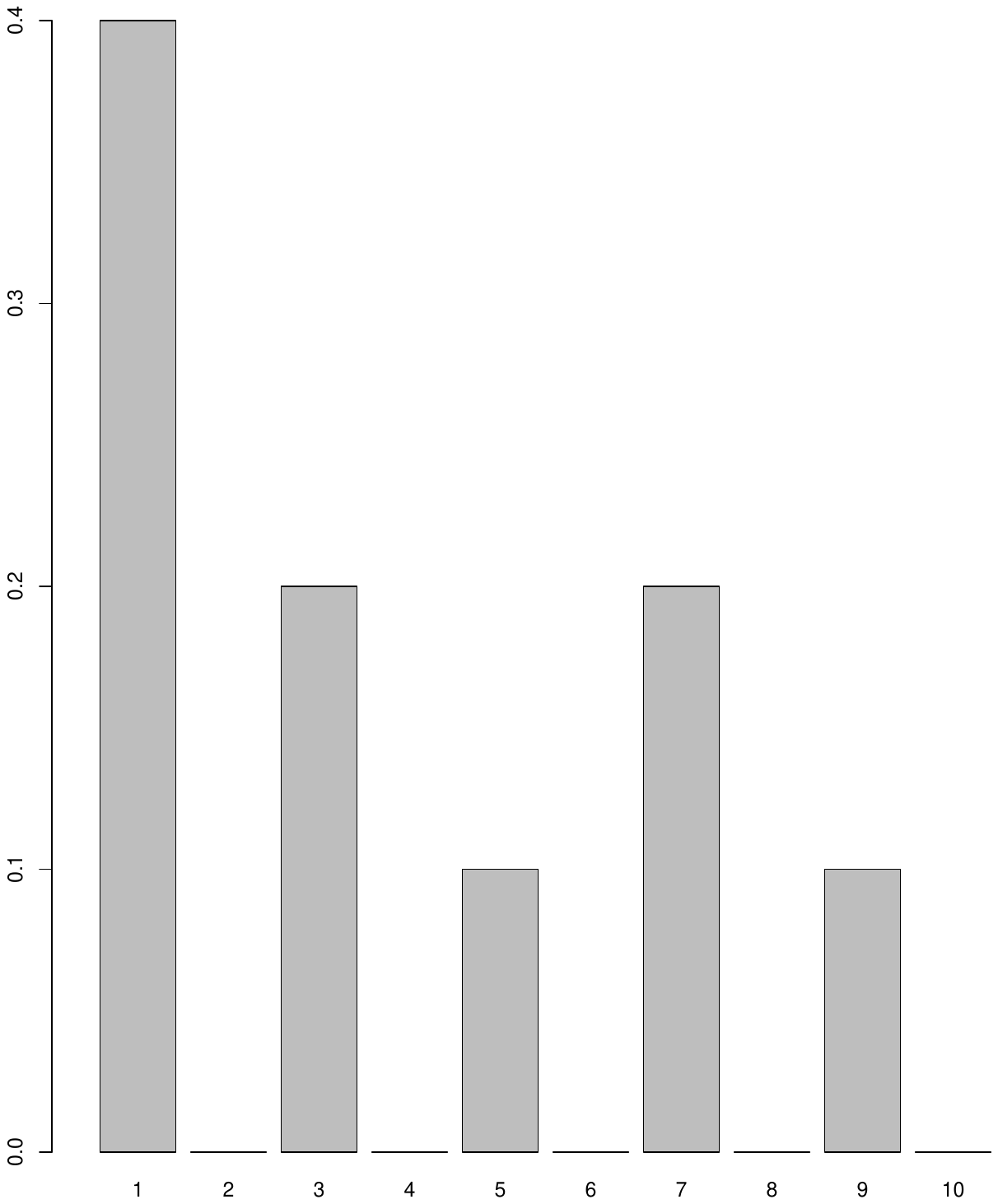}\label{fig:unsmoothed-histogram}}\quad
  \subfloat[Smoothing with $h=1$]{\includegraphics[width=0.25\textwidth]{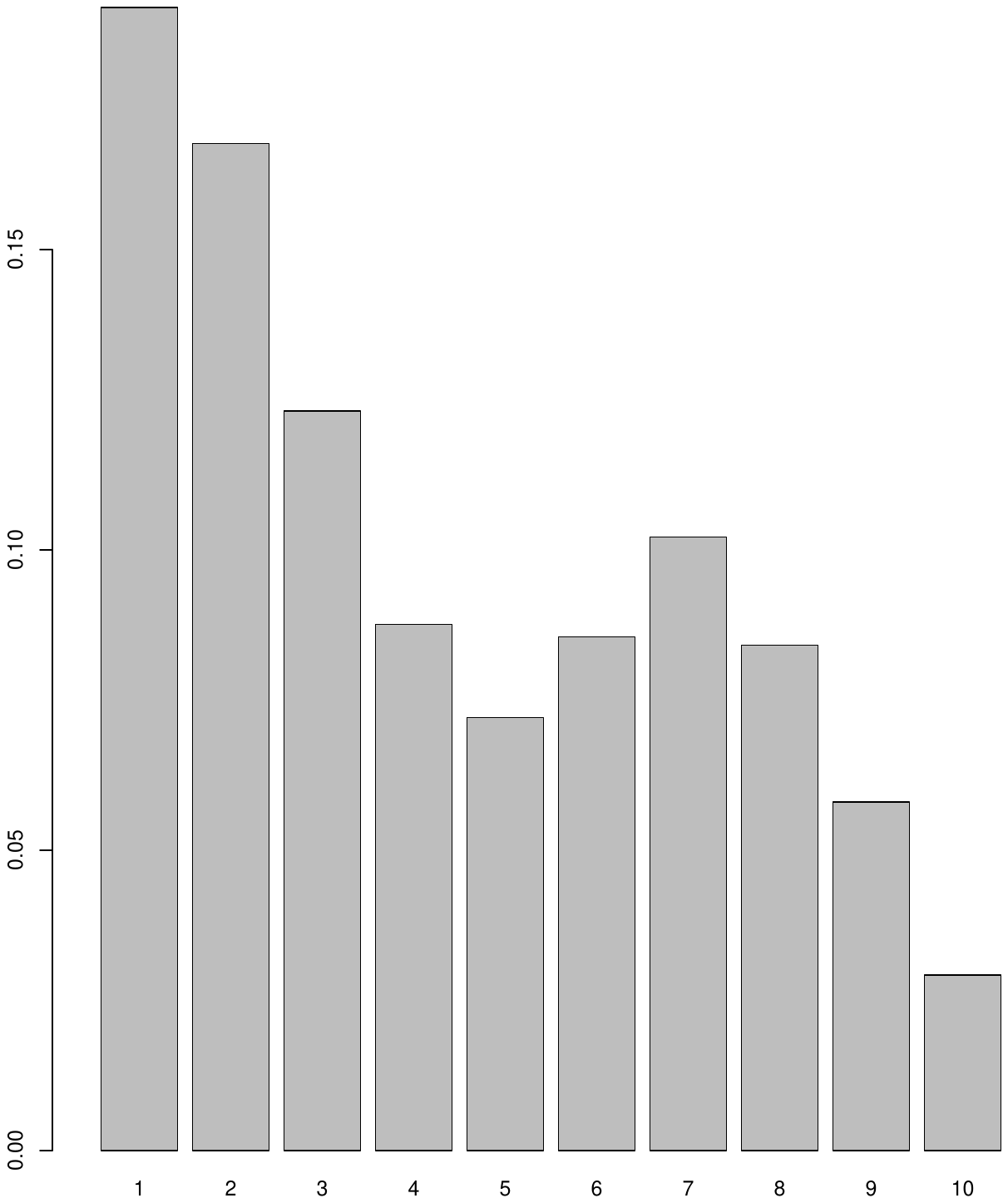}\label{fig:unsmoothed-histogram}}\quad
  \subfloat[Smoothing with $h=0.25$]{\includegraphics[width=0.25\textwidth]{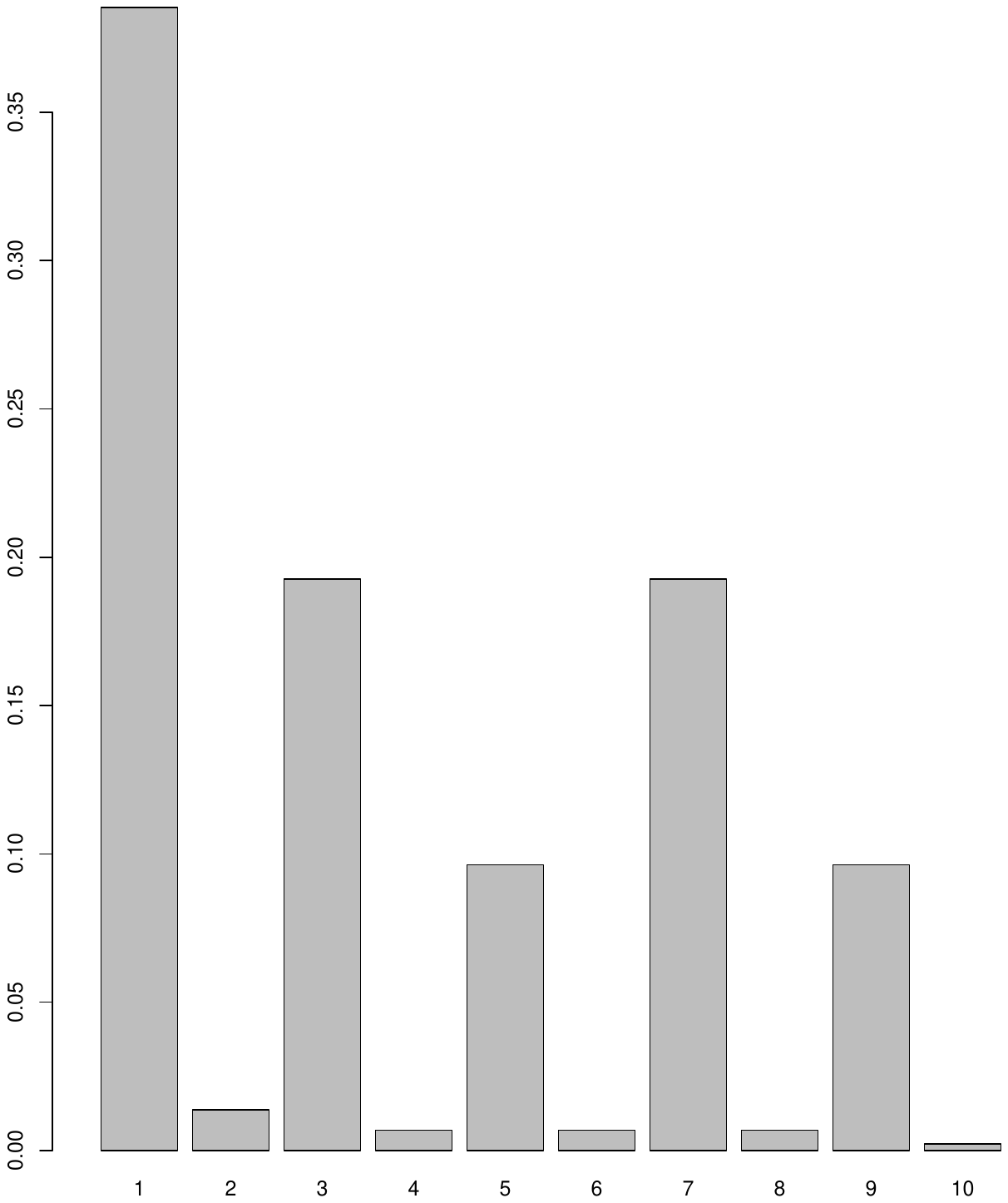}\label{fig:unsmoothed-histogram}}
  \caption{Effect of smoothing the histogram to fill empty categories and extend the support}\label{fig:smoothing-example}
\end{figure}

Of course, it will not be feasible to ask experts to provide entire
probability distributions over a scoring scale, but the smoothing of the
empirical distribution by convolution with $K_h$ achieves the same thing.
Viewing $K_h$ as a set of weights associated with all possibilities around
the mean value $x$, the expert assessment $x\in\Z$ receives the maximal
weight (as being the modal value of $K_h$), whereas all other possibilities
receive a nonzero weight that decays with the distance to the value $x$.
Graphically, this process thus ``fills'' all empty bins in the empirical
histogram (say, if a category has never been assigned during the expert
survey), and additionally extends the distribution up to the full and
identical support for all loss distributions. Hence, besides satisfying the
technical constraints imposed, the smoothing naturally accounts for the
uncertainty in the assessment; the lower the parameter $h$ is, the more
confident we are in the assessment (i.e., the closer the smoothed
approximation $\hat F_{n,h}$ approaches $\hat F_n$). Figure
\ref{fig:smoothing-example} displays an example of the original empirical
histogram (Figure \ref{fig:unsmoothed-histogram}), and its smoothed version
with $h=1$, having its gaps filled in by the superposition of smoothing
kernels (expressing the uncertainty around the modal values given in Figure
\ref{fig:unsmoothed-histogram}), and another smoothed version of the
histogram with a much smaller bandwidth parameter $h=0.25$. This plot
visually illustrates (confirms) the formal claim of convergence as stated
previously.

The quality of the kernel density estimation, as in general with this
nonparameteric method, is the art of choosing the bandwidth parameter $h$.
Various rules of thumb (e.g., Silverman's rule \cite{Silverman1998}) or
cross-validation techniques may be applied. Since there is no generally
``best'' way to choose this parameter depending on the information at hand,
we refer to the literature on nonparametric statistics for concrete methods
\cite{Wasserman2007,Tsybakov2009,Devroye1985}.


\section{Risk Prioritization}\label{sec:risk-priorization}
\begin{figure}
  \centering
  \includegraphics[scale=0.7]{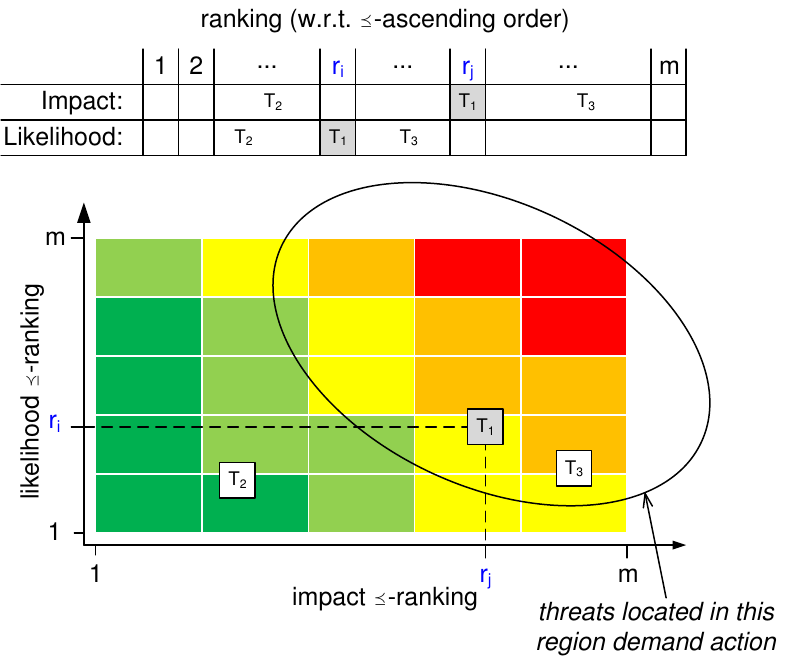}
  \caption{Risk Matrix using $\preceq$-Orderings}\label{fig:risk-matrix-hyrim}
\end{figure}

Let us assume that a total of m threats $T_1, \ldots ,T_m$ has been
identified. If, say $n$, experts provide their input on the impact and
likelihoods (both artefacts obtained by surveys as outlined in section
\ref{sec:expert-surveys}), then it is a simple matter to compile an empirical
histogram (a distribution $F_I$) for the impact and another empirical
histogram (a distribution $F_L$) for the likelihood. In doing so for every
threat $T_1, \ldots ,T_m$, we end up with $2m$ such distributions, which we
denote here as $F_(k,I),F_(k,L)$ for the $k$-th threat $T_k$ in the list. To
resemble the risk matrix familiar from the standard process of risk
management, we can use the $\preceq$-ordering on the distributions (remember
that it is a total order), to separately sort the threats $T_1,\ldots,T_m$ in
ascending $\preceq$-order according to their impact and likelihood. This
gives two lists, in which each threat gets a rank assigned among the total of
$m$ threats, say the $k$-th threat has rank $r_i$ on the impact ranking and
position $r_j$ on the likelihood ranking. The risk matrix familiar from the
ISO processes (cf. Figure \ref{fig:risk-matrix}) is then re-established by
placing each threat into a 2D-coordinate system with the $x$-coordinate being
the rank on the impact scale and the $y$-coordinate being the rank on the
likelihood scale. This puts each threat to a particular position on the grid,
and we may then proceed ``as usual'' by coloring the area as we would do for
a normal risk matrix. Likewise, the ``critical region'' can be defined as the
area inside which threats fall that need to be addressed by the subsequent
risk management process. Figure \ref{fig:risk-matrix-hyrim} displays an
example.

The main difference to the standard risk matrix is herein twofold:
\begin{enumerate}
  \item While a standard risk matrix has axes scaled in categories, the so
      adapted risk matrix has axes scaled in ranks that go from 1 to the
      total number $m$ of threats.
  \item The consensus problem of assigning a threat a single quantification
      in terms of likelihood and impact is avoided. Since many experts can
      utter disagreeing opinions, all of which go into the impact- and
      likelihood-distributions $F_I$ and $F_L$ with equal importance. The
      total ordering $\preceq$ then assures somewhat like a
      ``base-democratic'' ranking, since one threat outranks the other in
      the $\preceq$-ordering, if more people classify the impact,
      respectively the likelihood, as high (see \cite{Rass2015b}~
      for the full detailed effect of this ordering).
\end{enumerate}

If the assessment considers multiple criteria, say, if the impact is not only
measured in money but also in reputation, then the respective assessments are
made separately (not necessarily independently) from one another. For the
example of budget-impact and reputation impact, we could think of the one
distribution $F_I$ as being replaced by two new distributions $F_{BI}$ and
$F_{RI}$ (for budget and reputation). Likewise, if two threats have
assessments in these terms, given as $F_{(BI,1)},F_{(BI,2)}$ and
$F_{(RI,1)},F_{(RI,2)}$, then we ought to rank the two in terms of two
criteria. A canonical way of doing that is offered by defining a game with
two goals (``budget'' and ``impact''), and using one (dummy) strategy for
player 1, and letting the two threats being the strategies for the opponent
(player 2). It is, however, important to stress that we \emph{cannot}
directly go ahead and set up a multi-criteria $(1\times n)$-game to get the
most severe threat via an equilibrium, since the gameplay is defined to be
among $n+1$ players, where each opponent (corresponding to a goal) plays
independently of all others. The equilibrium would then return a worst-case
threat identified \emph{per goal}, which is possibly not what we seek here
for a multi-criteria risk optimization. The theory, however, remains
applicable when the game is reduced to a two-player $(1\times n)$ game with
scalar(ized) payoffs per player. In that case, the equilibrium is necessarily
pure and indicates the worst case threat for player 1. The \texttt{R} package
implements this method in a designated procedure \texttt{preference}:

\begin{verbatim}
# ranking if there is only one goal (say, FBI1, vs. FBI2)
> preference(FBI1, FBI2)
2 # this means that the second parameter (FBI1) is preferable

# ranking, if there are multiple goals (of equal importance)
> preference(list(FBI1, FRI1), list(FBI2, FRI2))

# Let us assign twice as high priority to the "reputation"
# goal by supplying the parameter weight=c(1,2), i.e. "goal" has
# priority 1, and "reputation" has the (double) priority 2.
> preference(list(FBI1, FRI1), list(FBI2, FRI2), weights=c(1,2))
\end{verbatim}
The results directly gives the desired ranking by telling that either the
first (output ``1'') or second (output ``2'') threat is \emph{less severe};
if the function returns zero, then the distributions are identical and the
decision is indifferent. For three or more threats, the procedure can be
repeated pairwise to rank a whole set of threats in the way as described
above.

\section{Examples of Game-Theoretic Modeling}

\subsection{Modeling APTs}\label{sec:apt-modeling}
The investigation of many examples of advanced persistent threats (APTs)
reported in the past reveal a common structure underlying an APT, even though
the details thereof may be highly different. Quoting the taxonomy of
\cite{DellSecureWorks2014}, an APT roughly proceeds along the following
steps:
\begin{enumerate}
\item Initial infection:
    \begin{enumerate}
      \item reconnaissance: information gathering,
      \item development: design of a made to measure malware,
      \item weaponization: preparing the trojan and droppers,
      \item delivery: transmission into the victim infrastructure, e.g.,
          by a phishing email, or similar.
    \end{enumerate}
\item Learning and propagation: repeated sequence of
    \begin{enumerate}
      \item exploitation: to get deeper into the system
      \item installation: to leave artifacts and backdoors for an easy
          return later, and destroy footprints of the attack
    \end{enumerate}
\item Damage:
    \begin{enumerate}
      \item command and control: interaction with the victim system's
          compromised resources via previously left artifacts
      \item actions on the target: causing the actual damage
    \end{enumerate}
\end{enumerate}

The specific actions taken in each of these phases depend on the target
infrastructure and no general description is possible due to the diversity of
such infrastructures. However, specific examples can be ``read off'' reported
prominent incidents, such as including Stuxnet \cite{Farwell2011}, Duqu
\cite{Bencsath2012}, Flame \cite{Munro2012}, and Aurora \cite{Kurtz2010}. A
common element in all these appears to be the human factor, which we discuss
separately in section \ref{sec:human-element}.

Game theory has been applied in various ways to model attacks and cyber risks
\cite{Manshaei2013,Dijk2013,Zhu2013} and proposals (independent of game
theory) include moving target defenses \cite{Jajodia2011,Jajodia2013}, trust
mechanisms \cite{Uzal2013}, and defense-in-depth techniques
\cite{Pawlick2015}. Here, we will review a recent proposal to model APTs as a
sequence of games, each of which is tailored to the specific nature and
details of the respective phase.

That is, an APT usually begins with a harmlessly looking email or ``lost''
USB stick, on which malware enters the system. Alas, nowadays attackers have
joint forces into an entire illegal business sector covering the entire
supply chain of cyber crime in a spectrum of independently offered services
(often referred to as ``cybercrime as a service'' in alignment to cloud
computing terminology). This means that the person identifying a weakness is
not necessarily the same who is writing the exploit for it. Likewise, the
author of the exploit kit is not necessarily applying it anywhere, but merely
sells it to the actual attacker. Similarly, infections with malware are not
automatically intended to cause immediate damage. The term
``botnet-as-a-service'' describes the business model of infecting a large
number of machines and offering to deploy malware on these zombies upon
request and for a (smaller or larger) fee. In this way, the attacker can
simply ``buy'' access to a potentially large number of infected machines
instantly. For the victim, this has the unpleasant effect of the infection
remaining stealthy and inactive, until some time later, when the outbreaks
causes noticeable damage. At this time, however, it is most likely that no
connection between a past email and the current incident is recognized. The
exact time window between an infection and its activity is also dependent on
the technical countermeasures adopted in the specific company. For example,
signature-based malware recognition (like classical anti-virus software used
to work) may take a couple of days until the malware is recognized as such
and the respective signature is shipped with the next update. On the
contrary, cloud based malware detection that is based on recognizing a huge
lot of ``identical'' email (attachments) suddenly flowing through the
internet can have a much shorter time (even a few hours) until a malware is
suspected and classified. So, the time window between the infection and the
exploit may, in some cases, be closed quite fast (by a good malware
recognition system).

To model a so-structured APT in terms of games, the first step is listing all
potential ways into the system, such as include (but are not limited to):
\begin{itemize}
  \item (spear) phishing, whaling,
  \item waterholing,
  \item tailgating,
  \item etc.
\end{itemize}
The entirety of these possibilities makes up the set of strategies $PS_2$ for
the attacker. Typically, an entry in $PS_2$ is thus not constrained to have a
specific form or structure (in particular, an element $x\in PS_2$ is not a
purely mathematical object but rather a \emph{description} of how an attack
would be launched according to this particular strategy $x$).

In a different view, $PS_2$ can be considered as a list of threats, to which
corresponding countermeasures can be defined (appealing to catalogues found
in standards like ISO 27000, etc.). The list of countermeasures is $PS_1$.

Note that $PS_1$ and $PS_2$ should be defined bearing in mind the steps and
structure of the first phase of the APT, which comprises information
gathering, development, weaponization and delivery. Consequently, example
defense measures in $PS_1$ may relate to periodic (and random)
re-configurations of the system in order to thwart the attacker's respective
next steps, either by invalidating so-far collected information or by
removing malware by coincidence, say if a computer is reinstalled or
privileges of the victim user have changed (were revoked).

In any case, let the game describing the initial infection be a matrix game
$G_1$ (where the subscript is a reminder of this being the \emph{first} phase
of the APT). It appears fair to let this be a matrix game (static and
repeated), since if the game is expectedly repeated either in case of an
attack failed, or to infect further parts of the system, if an attack was
successful.

Let the equilibrium computed for the infection game $G_1$ be the pair of
distributions $(F_1^*, F_2^*$), where the optimal defense against the initial
infection is the distribution $F_1^*$ (the distribution $F_2^*$, more
specifically, its probability mass function can be taken as a non-unique
indicator set of neuralgic points in the infrastructure).

To model the next phase of learning and propagation, we divide the (physical)
infrastructure in \emph{stages} that the adversary needs to reach one by one
in order to get to the inner target asset. If the infrastructure is a network
graph $G(V,E)$, then the target asset may be a(ny) fixed node $v_0\in V$, and
the \emph{$i$-th stage} can (but does not need to) be \emph{defined as the
set of nodes at distance $i$ to the node $v_0$}. That is, the stages are
``concentric circles'' around the asset, and on each stage, the adversary
plays a (different) game to get to the next stage. A single such stage is
penetrated by the aforementioned steps of \emph{exploitation} and
\emph{installation}, continuing with exploitation again on the next stage
(only in a different setting there). This modeling is practically supported
by topological vulnerability analysis (TVA), which can cook up an attack
graph for an infrastructure. In the following, we will illustrate our
thoughts based on Example \ref{exa:attack-graph}.

\begin{exa}[based on \cite{Singhal2011}; see also \cite{Rass&Konig&Schauer2017}]\label{exa:attack-graph}
Consider a system as shown in Figure \ref{fig:apt-example-tva}, composed from
three devices, with various ports opened and services enabled.
\begin{figure}[t!]
  \centering
\subfloat[Example Infrastructure \cite{Singhal2011}]{\includegraphics[scale=0.7]{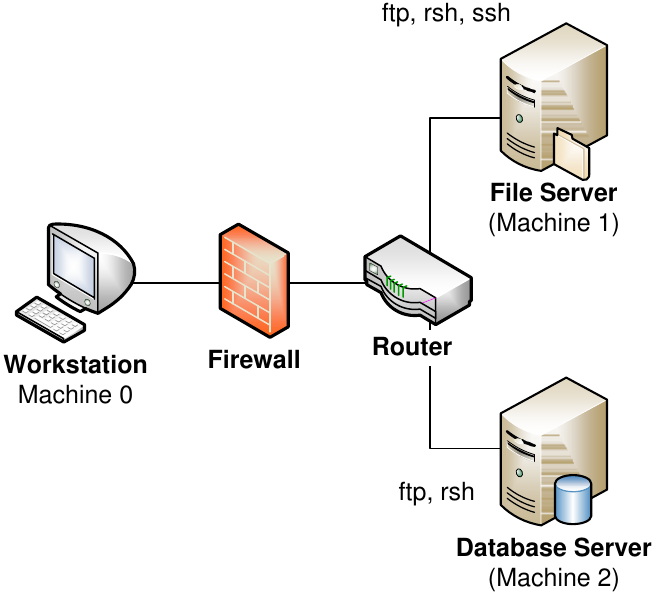}\label{fig:example-infrastructure}}
\subfloat[Example Attack Graph \cite{Singhal2011}]{\includegraphics[width=0.5\textwidth]{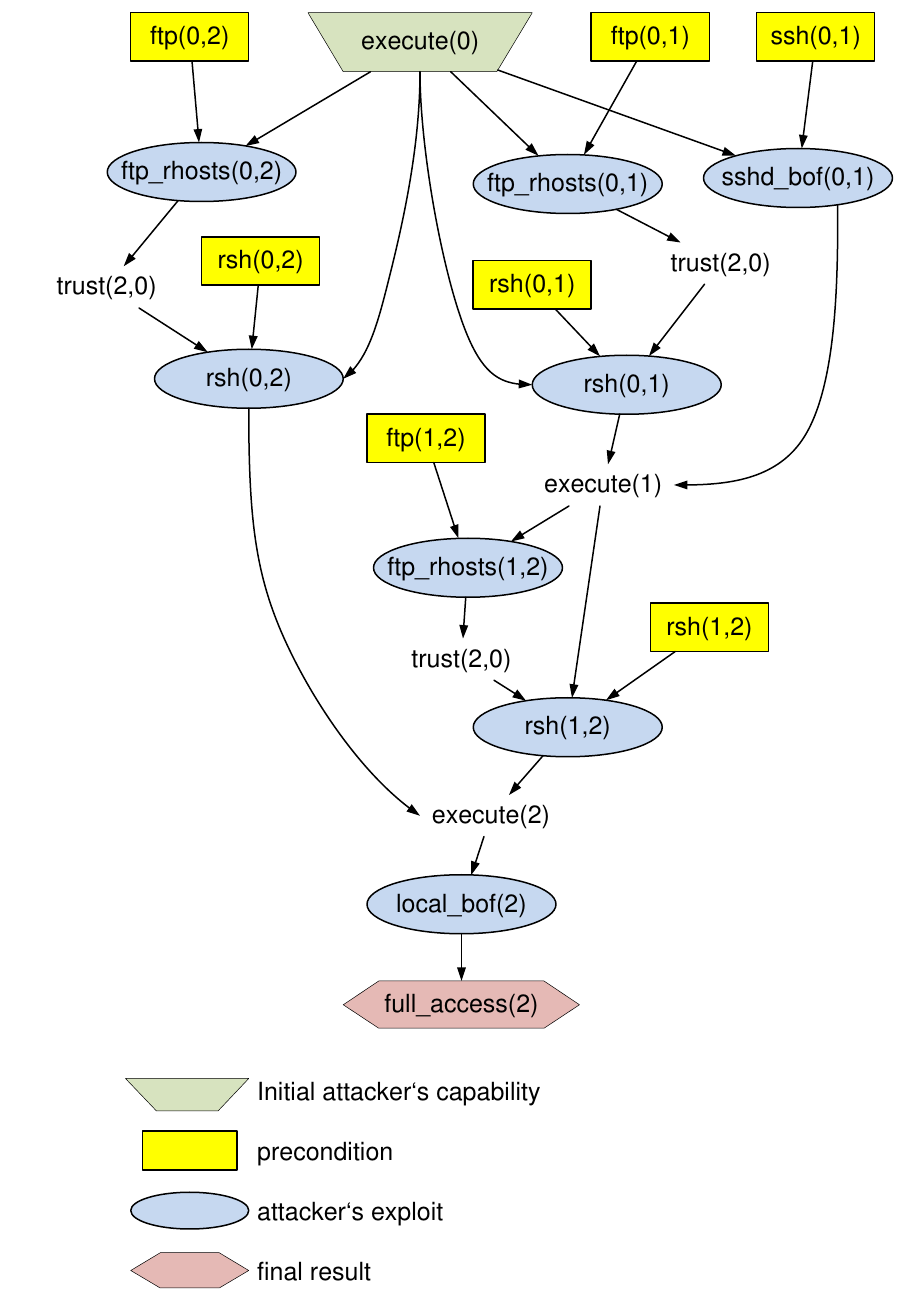}\label{fig:apt-example-tva}}
\caption{Topological Vulnerability Analysis Example}\label{fig:tva-example}
\end{figure}
Based on this information, the attacker can consider several exploits, such
as:
\begin{itemize}
  \item FTP- or RSH-connections from a node \texttt{x} to a remote host
      \texttt{y}, hereafter denoted as \texttt{ftp\_rhosts(x,y)}, and
      \texttt{rsh(x,y)}, respectively.
  \item a secure shell buffer overflow at node \texttt{y}, remotely
      initiated from node \texttt{x}, hereafter denoted as
      \texttt{sshd\_bof(x,y)}.
  \item local buffer overflows in node \texttt{x}, hereafter denoted as
      \texttt{local\_bof(x)}.
\end{itemize}
Each of these may establish a trust relation between two nodes \texttt{x} and
\texttt{y}, which we denote as \texttt{trust(x,y)}. The list of attack
strategies to penetrate all the stages until \texttt{full\_access} to machine
2 is a matter of plain path enumeration in the attack graph, whose results
are shown in table \ref{tbl:as1}.

\begin{table}[t!]
  \centering
  \scriptsize
  \caption{APT scenarios (adversary's action set $PS_2$, based on Figure \ref{fig:tva-example})}\label{tbl:as2}
  \begin{tabular}{|l|p{0.8\textwidth}|}
    \hline
    1 & \texttt{execute(0)} $\to$ \texttt{ftp\_rhosts(0,1)} $\to$ \texttt{rsh(0,1)} $\to$ \texttt{ftp\_rhosts(1,2)} $\to$ \texttt{sshd\_bof(0,1)} $\to$ \texttt{rsh(1,2)} $\to$ \texttt{local\_bof(2)} $\to$ \texttt{full\_access(2)} \\\hline
    2 & \texttt{execute(0)} $\to$ \texttt{ftp\_rhosts(0,1)} $\to$ \texttt{rsh(0,1)} $\to$ \texttt{rsh(1,2)} $\to$ \texttt{local\_bof(2)} $\to$ \texttt{full\_access(2)}\\\hline
    3 & \texttt{execute(0)} $\to$ \texttt{ftp\_rhosts(0,2)} $\to$ \texttt{rsh(0,2)} $\to$ \texttt{local\_bof(2)} $\to$ \texttt{full\_access(2)} \\\hline
    4 & \texttt{execute(0)} $\to$ \texttt{rsh(0,1)} $\to$ \texttt{ftp\_rhosts(1,2)} $\to$ \texttt{sshd\_bof(0,1)} $\to$ \texttt{rsh(1,2)} $\to$ \texttt{local\_bof(2)} $\to$ \texttt{full\_access(2)} \\\hline
    5 & \texttt{execute(0)} $\to$ \texttt{rsh(0,1)} $\to$ \texttt{rsh(1,2)} $\to$ \texttt{local\_bof(2)} $\to$ \texttt{full\_access(2)} \\\hline
    6 & \texttt{execute(0)} $\to$ \texttt{rsh(0,2)} $\to$ \texttt{local\_bof(2)} $\to$ \texttt{full\_access(2)} \\\hline
    7 & \texttt{execute(0)} $\to$ \texttt{sshd\_bof(0,1)} $\to$ \texttt{ftp\_rhosts(1,2)} $\to$ \texttt{rsh(0,1)} $\to$ \texttt{rsh(1,2)} $\to$ \texttt{local\_bof(2)} $\to$ \texttt{full\_access(2)} \\\hline
    8 & \texttt{execute(0)} $\to$ \texttt{sshd\_bof(0,1)} $\to$ \texttt{rsh(1,2)} $\to$ \texttt{local\_bof(2)} $\to$ \texttt{full\_access(2)} \\
    \hline
  \end{tabular}
\end{table}

A selection of respective countermeasures is given in table \ref{tbl:as1}.

\begin{table}
  \centering
  \scriptsize
  \caption{Security controls (selection) -- subset of $PS_1$}\label{tbl:as1}
  \begin{tabular}{|p{0.2\textwidth}|p{0.7\textwidth}|}
    \hline
    countermeasure & comment \\\hline\hline
    deactivation of services (FTP, RSH, SSH) & these may not be permanently disabled, but could be temporarily
      turned off or be requested on demand (provided that either is
      feasible in the organizational structure and its workflows) \\ \hline
    software patches & this may catch known vulnerabilities (but not necessarily all of
      them), but can be done only if a patch is currently available \\ \hline
    reinstalling entire machines & this surely removes all unknown malware but comes at the cost of
      a temporary outage of a machine (thus, causing potential trouble with
      the overall system services) \\  \hline
    organizational precautions & for example, repeated security training for the employees. These may
      also have only a temporary effect, since the security awareness is
      raised during the training, but the effect decays over time, which
      makes a repetition of the training necessary to have a permanent
      effect. \\
    \hline
  \end{tabular}
\end{table}

The game to model the penetration can then be defined \emph{per stage} by
defining the machines 0, 1 and 2 as stages, where machine 2 is the inner
assert (node $v_0$ in our previous wording), and a stage is the game played
to establish a trust relation between a machine at distance $i$ and one at
distance $i-1$ to machine 2.

In the (quite simple) infrastructure of Figure
\ref{fig:example-infrastructure}, the stages would thus be:
\begin{itemize}
  \item stage 1 (distance 1 to machine 2): \{router\}
  \item stage 2 (distance 2 to machine 2): \{file server (machine 1),
      firewall\}
  \item stage 3 (distance 3 to machine 2): \{workstation (machine 0)\}
\end{itemize}
The game played at stage 3, accordingly, has strategies equal to all exploits
that can be mounted on the workstation (machine 0), which can be read off the
TVA attack tree (Figure \ref{fig:apt-example-tva}) as $PS_2
=\{$\texttt{ftp\_rhosts(0,1)}, \texttt{ftp\_rhosts(0,1)},
\texttt{sshd\_bot(0,1)}\}. The corresponding set $PS_1$ comprises all
countermeasures that can be implemented (such as virus checks, temporary
disabling of services, but also non-technical ones like repeated security
training, etc.). The strategy spaces $PS_1$ and $PS_2$ identified in this way
then define the shape of the respective 3rd stage game $G_3$, whose payoff
structure is to be defined following the procedure outlined in section
\ref{sec:expert-surveys}; Figure \ref{fig:workflow}.

The games for the other stages are constructed analogously.
\end{exa}

Given a game for each stage, we can connect them into an overall model for
phase two of the APT by adopting a high level perspective. In each stage, the
adversary has basically two options, which are:
\begin{enumerate}
  \item \emph{penetrate}: this means launching an attack as identified
      based on previously gathered information (cf. example
      \ref{exa:attack-graph}), or
  \item stay, to collect more information while remaining stealthy $\to$
      \emph{learning}.
\end{enumerate}
Both options can be modeled using their own (distinct) game models, and the
connection between the games over all stages is established by considering
that in the $n$-th stage game (whether we go for penetrating or staying), the
outcome is one of the following:
\begin{itemize}
  \item if the attacker decides to penetrate, then
    \begin{itemize}
      \item it may succeed, in which case it enters game $G_{n-1}$, or
      \item it may fail, in which case it has to repeat game $G_n$ once
          more.
    \end{itemize}
  \item if the attacker decides to stay, then
    \begin{itemize}
      \item it may succeed and gain further information, but this leaves
          the attacker in this stage $n$,
      \item it may fail, in which case the overall game terminates, and
          the entire investment of the attacker is lost (we model this as
          a negative gain for the attacker).
    \end{itemize}
\end{itemize}

Likewise, the defender has the option of defending (without any guarantee of
the defense being successful), or not defending. The latter strategy is
implicitly played whenever the defender ``pauses'' in its defense, say, if
the security guard is being sent elsewhere and misses the attack by this
unfortunate coincidence. In any case, ``do not defend'' is never a dominating
strategy and occurs only due to resource limitations and the inability to
defend everywhere at all times against everything.

Writing $I(n)$ for the payoff (to the adversary) in the $n$-th stage, the
above modeling yields a $2\times 2$ zero-sum matrix game structured as shown
in Figure \ref{fig:sequential-apt-game} (cf. \cite{Rass&Zhu2016}). In this
model, we have additional quantities $p(n)$ and $q(n)$ that depend on the
current stage, and quantify the likelihood for each action to be successful.
These values can be either defined directly, or themselves be derived as
saddle point values of classical 0-1-valued matrix games in each stage (in
this case, the risk assessment is done in binary terms only asking the expert
for whether or not the attack will be successful).

\begin{figure}
\small
\centering
\begin{tabular}{|c|c|c|}
  \hline
  \backslashbox{defender}{attacker} & penetrate & stay \\\hline
  defend & $p(n)\cdot I(n-1) + (1-p(n))\cdot I(n)$ & $q(n)\cdot I(n) + (1-q(n))\cdot (-I(n))$ \\\hline
  do not defend & $I(n-1)$ & $I(n)$ \\
  \hline
\end{tabular}
\caption{Sequential 2-Player Game Model for Advanced Persistent Threats}\label{fig:sequential-apt-game}
\end{figure}

It is reasonable to assume a circular structure in this game (cf. also
\cite{Avenhaus2002}), which amounts to a single unique equilibrium, for the
following reasons:
\begin{itemize}
  \item if ``defend'' is a dominating (row) strategy, then either ``stay''
      or ``penetrate'' will complete this into an equilibrium, and there is
      nothing to be optimized by game theory here on this (higher) level.
  \item if ``do not defend'' is a dominating (row) strategy, then there is
      no need to do any active security here, as the infrastructure is
      already secure anyway.
  \item if ``penetrate'' is a dominating (column) strategy, then the
      obvious best choice is to defend, which again degenerates the game
      into a trivial matter.
  \item if ``stay'' would be a dominating (column) strategy, then there is
      no need to defend anything, since the attacker is not trying to get
      to the asset anyway.
\end{itemize}
In any case, it is a simple (yet laborious) matter of working out the game
matrix for each stage, and determine an equilibrium for it (say, using
fictitious play as described in \cite{Rass2015b}). 
The particular equilibrium obtained for the $n$-th stage is the distribution
$I(n)$, telling us the distribution of damage in that stage, accompanied by
optimal defenses in this stage (obtained from the $2\times 2$ game and the
inner sub-games played for penetration and to collect information during a
stay).

Concluding the idea, the decomposition of the learning-and-penetration phase
in the APT life cycle into stages and corresponding games played therein
delivers an in-depth defense action plan (individual randomized defense
actions being taken on each part in the infrastructure), as well as a
higher-level risk assessment $I(1), I(2), \ldots$ that refers to each stage.
The distribution $I(n)$ for the $n$-th such stage then indicates the
likelihood of damage suffered at the $n$-th ``protective layer'' around the
asset of interest.

The final game modeling the Damage-phase of the APT can then be defined
similar to the initial infection game $G_1$, as the attacker may simply try
and retry causing damage. The process and steps to identify possible actions
and countermeasures must again be supported by special purpose (and
application specific) tools and expertise, but the game theoretic treatment
and computation of risk metrics remains the same. Thus, we will not repeat
the details here.

\paragraph{A Static Game Model for the Penetration Phase}
In a more simplified view towards a substitute of the sequential phase two
game, a static game model can be considered as an alternative (being easier
to model and more efficient to analyze computationally). This simplification
is bought at the cost of getting a more coarse-grained model, since the
attack and defense strategies are defined more ``high-level'' and not
specific for each stage in the graph representation (of the infrastructure or
the attack tree).

As before, we can think of the attacker working its way through the stages,
while occasionally being sent back or kicked out if a security officer
(perhaps unknowingly) closes some of the backdoors established previously.
The pure strategy set $PS_1$ for the defender is thus the set of all
nodes/components in the system on which spot checks (e.g., malware scans,
configuration changes, updates, patches, etc.) can be done. Different to the
per-stage game modeling from before, the defenses now correspond to
high-level attack strategies $PS_2$ that describe only techniques but are not
specifically tailored to a particular machine. That is, $PS_2$ would be
composed from a more generic set of threats like
\begin{itemize}
  \item buffer overflow exploits,
  \item cross-site scripting,
  \item code injections,
  \item etc.,
\end{itemize}
but unlike in Example \ref{exa:attack-graph}, these attacks would not be
considered on specific machines. Rather, the attack scenario is described as
a general code injection that may be tried on any machine in the network (if
possible). The description of the attack/defense scenario, as well as the
possibilities to consider for a risk assessment may be more complex in this
kind of modeling, as Example \ref{exa:simple-modeling} shall illustrate.

\begin{exa}\label{exa:simple-modeling}

The expert survey would ask something like:
\begin{quote}
What is the expected damage if an attacker attempts a code injection attack
in our infrastructure, while machine 1 (see Figure
\ref{fig:example-infrastructure}) is being re-installed at roughly the same
time.
\end{quote}
To answer this question towards a loss estimate, the expert may consider the
following aspects:
\begin{itemize}
  \item Which machines could be vulnerable to such an attack (say, which of
      them are accessible by a web interface or maybe have an insufficient
      patch level, etc.)
  \item Depending on which machine is attacked, the damage may be more or
      less (thus allowing the expert to utter multiple possibilities;
      substantiating the construction of a categorial loss distribution
      once again).
  \item Depending on where the spot check is done (say, on machine 1 in
      Example \ref{exa:attack-graph}), three cases are possible:
    \begin{itemize}
      \item Machine 1 has so far not been reached by the attacker (more
          precisely the attack path), so the reinstallation has no effect
          on the APT at this stage.
      \item Machine 1  is exactly the one currently targeted by the
          attacker. This may render the results of the APT learning phase
          on this machine useless, since the configuration has changed.
          Consequently, the attacker is sent back one stage and has to
          restart learning from here.
      \item Machine 1 has already been infected with malware, so that the
          adversary's backdoor path goes through it. In that case, the
          attacker's secret backdoor may be closed by the reinstallment
          and the attacker is again sent back to a previous stage.
    \end{itemize}
\end{itemize}
Since there is uncertainty on where the attack may be mounted, and also on
the current position of the adversary, the expert may utter several
possibilities in one of two ways:
\begin{itemize}
  \item Tell about a set of possibilities, expressing the likelihood for
      each of them individually. For example, the expert may say something
      about the most likely outcome, but in addition also say that more or
      less extreme results are possible with certain other likelihoods.
  \item Give a most likely outcome $r$, and express the uncertainty about
      it: this in fact corresponds to the aforementioned kernel density
      smoothing, since the uncertain variation around the most likely value
      (as expressed by the expert) may be well expressed by a bell-shaped
      curve centered around the value $r$. This is nothing else than a
      kernel density, where the uncertainty is ``quantified'' by the
      parameter $h$.
\end{itemize}
\end{exa}

\begin{figure}
  \centering
  \includegraphics[width=0.25\textwidth]{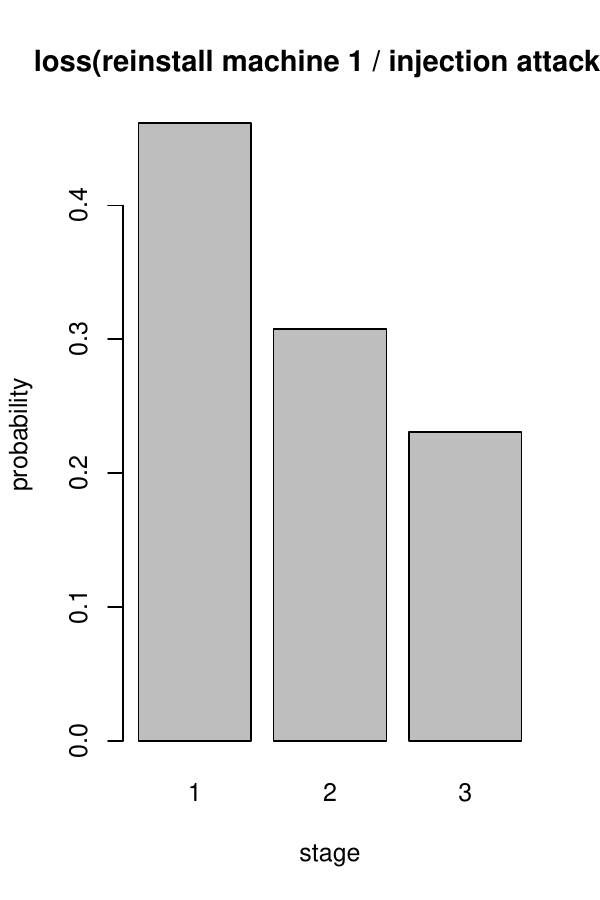}
  \caption{Example of loss distribution on stages (phase 2 of an APT)}\label{fig:stage-losses}
\end{figure}

In repeating the procedure exemplified in Example \ref{exa:simple-modeling}
for each (high-level) defense/attack scenario $(d_i,a_j)\in PS_1\times PS_2$,
we end up with a standard matrix game defined with probability-distribution
valued payoffs, which are now defined not on the risk categories, but on the
graph theoretic distances, or more specifically, the stages between the
initial infection and the inner asset. Picking up Example
\ref{exa:attack-graph}, there are three stages which the attacker has to
proceed through, and the loss distribution defined per defense/attack assigns
likelihoods to each of these stages; Figure \ref{fig:stage-losses} displays
an example. The game-theoretic optimization then goes for pushing the
probability mass towards ``more remote'' nodes in the network (or attack
graph). That is, we will try maximizing the distance between the attacker and
the asset, measured as the number of hops ($=$ stages in our wording) that
need to be taken to reach the goal. The stage numbering should, for
consistency with the minimization, be increasing from 1 up to the stage where
the asset is, i.e., stage 1 should comprise the outermost perimeter, followed
by the adjacent inner nodes, until stage $N>1$ being the set of final nodes
adjacent to the target asset.

\subsection{Modeling Social Engineering}\label{sec:human-element}

Social engineering is a good example of a case where the distribution-valued
game-theoretic framework perfectly fits. Since there is hardly any technical
countermeasure against such attacks, and the exploits are based on
psychological principles of human behavior, there is also an element of
``forgetting'' that makes security awareness decay over time. For this
reason, security training, information campaigns and similar must be repeated
from time to time, and there is never a guarantee that any of these
precautions has any or even durable effect. Both of these reasons render
matrix games with distribution-valued payoffs into a nice model, since:
\begin{itemize}
  \item the necessity of repeating social engineering awareness training
      corresponds with the modeling assumption on the matrix game to be
      repeated.
  \item the finiteness of matrix games corresponds to the limited set of
      possible countermeasures known against such attacks. That is, we
      cannot ask an employee just to become ``creative'' in how social
      engineering may be detected, and the best we can hope for is a
      feasibly small set of recommendations that a person can remember to
      avoid falling victim to social engineering.
  \item the inevitable element of human error renders the outcome of a
      social engineering scenario in any case up to randomness. Hence, our
      specification as a distribution-valued matrix game appears as a good
      fit, as it allows us to classify social engineering countermeasures
      as being ``variably effective'' (taking into account the differences
      in people's personality, daily mood, current workload, technical
      skills, time since the last security awareness training, and many
      more).
\end{itemize}

For a pure model of social engineering attack/defense scenarios, the
questionnaire outlined in section \ref{sec:expert-surveys} may be adapted to
poll people about their awareness against social engineering, describing the
particular attack scenario in the description of the question, and asking the
employee on how he/she would behave in this scenario. The results then
equally well compile into the sought loss distributions, if each answer in
the multiple-choice survey has a background association with a predefined
loss category (cf. section \ref{sec:loss-categories}).

A different approach to account for social engineering when it comes to loss
distributions is considering these attacks as methods to establish the
initial infection in an APT model as outlined in section
\ref{sec:apt-modeling}. Here, we would include social engineering techniques
in the attacker's strategy space $PS_2$, and the specification of the
respective success rates can be done using the same kind of survey as before.
The resulting game then already models the first phase of an APT infection,
in giving the probability for a social engineering attack to succeed under
the given security awareness campaign (which is a mixed strategy over $PS_1$
in the sense of repeated randomly selected trainings, information broadcast,
etc.).

The second phase of the APT, the penetration, can as well use social
engineering techniques to overcome barriers within the system. Suppose that a
subnetwork, for security reasons, does not have any physical or logical
connection to the outside or any other intranet within the company. Then a
malware can jump over this physical separation by a bring-your-own-device
incident. That is, if the malware gets into the system by someone connecting
a virulent USB stick to the inner separated network, the infection has
effectively overcome the logical separation. Even more, this scenario may
start from within the company's perimeter, since the infection of the USB
stick may indeed happen on an employees' computer, which has a connection to
the internet and got infected in the first phase.

The respective loss distributions associated with such an infection outbreak
can effectively be constructed by simulation, as is eloquently outlined in
\cite{Koenig2016a}. We leave the details aside here.

\section{Working with Game-Theoretic Risk Measures}
The description in the following is based on the previous explanations about
the model, and therefore only discusses practical matters of choosing the
parameters and hints on how to interpret the results. A description on how to
do the calculations with aid of the \texttt{R} statistical software suite is
given in section \ref{sec:tool-support}.

\subsection{Choosing the Cutoff Point}
\begin{figure}
  \centering
  \subfloat[Cutoff at $a=10$]{\includegraphics[height=0.4\textheight]{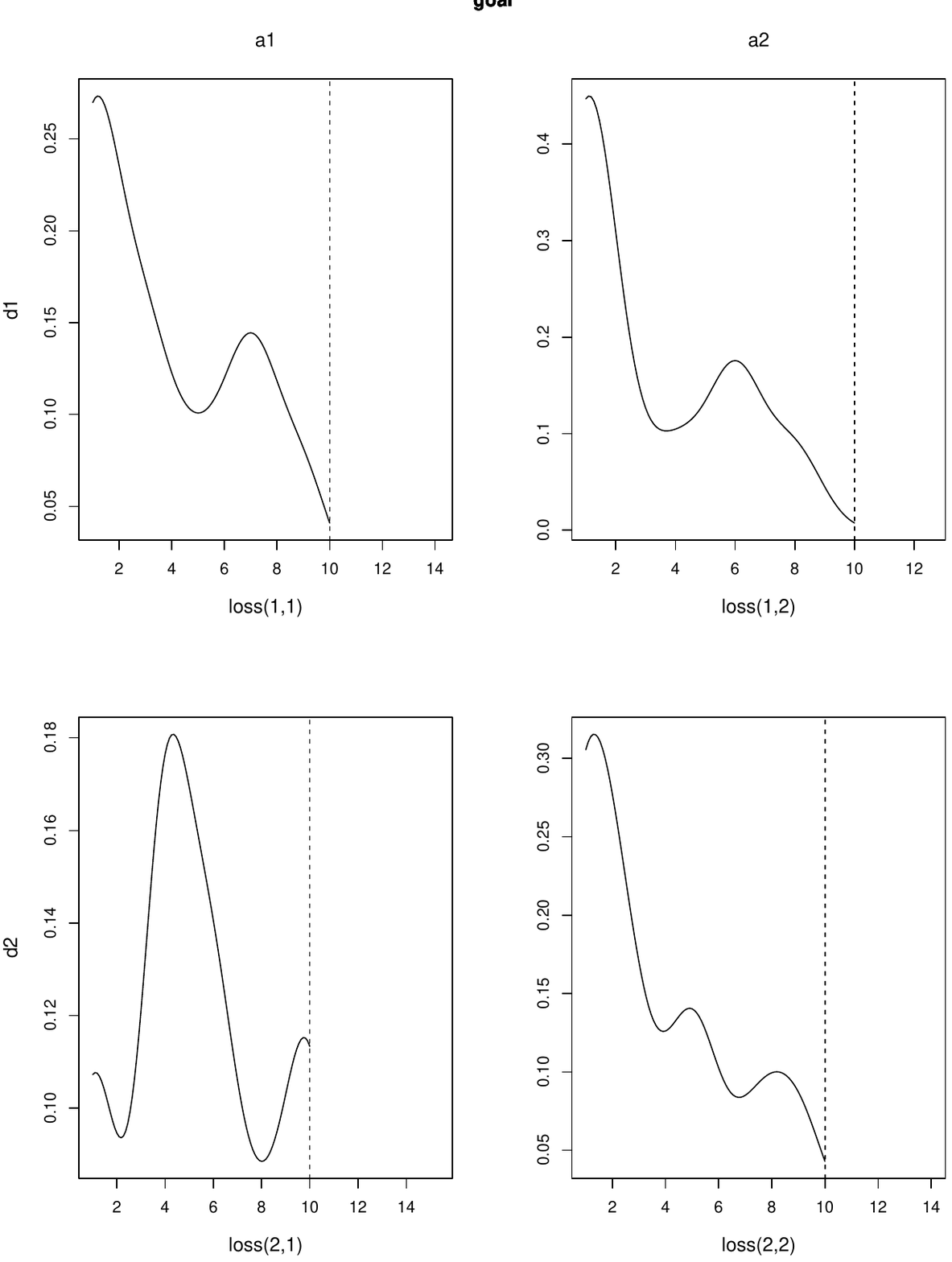}\label{fig:cutoff-10}}\\
  \subfloat[Cutoff at $a=6$]{\includegraphics[height=0.4\textheight]{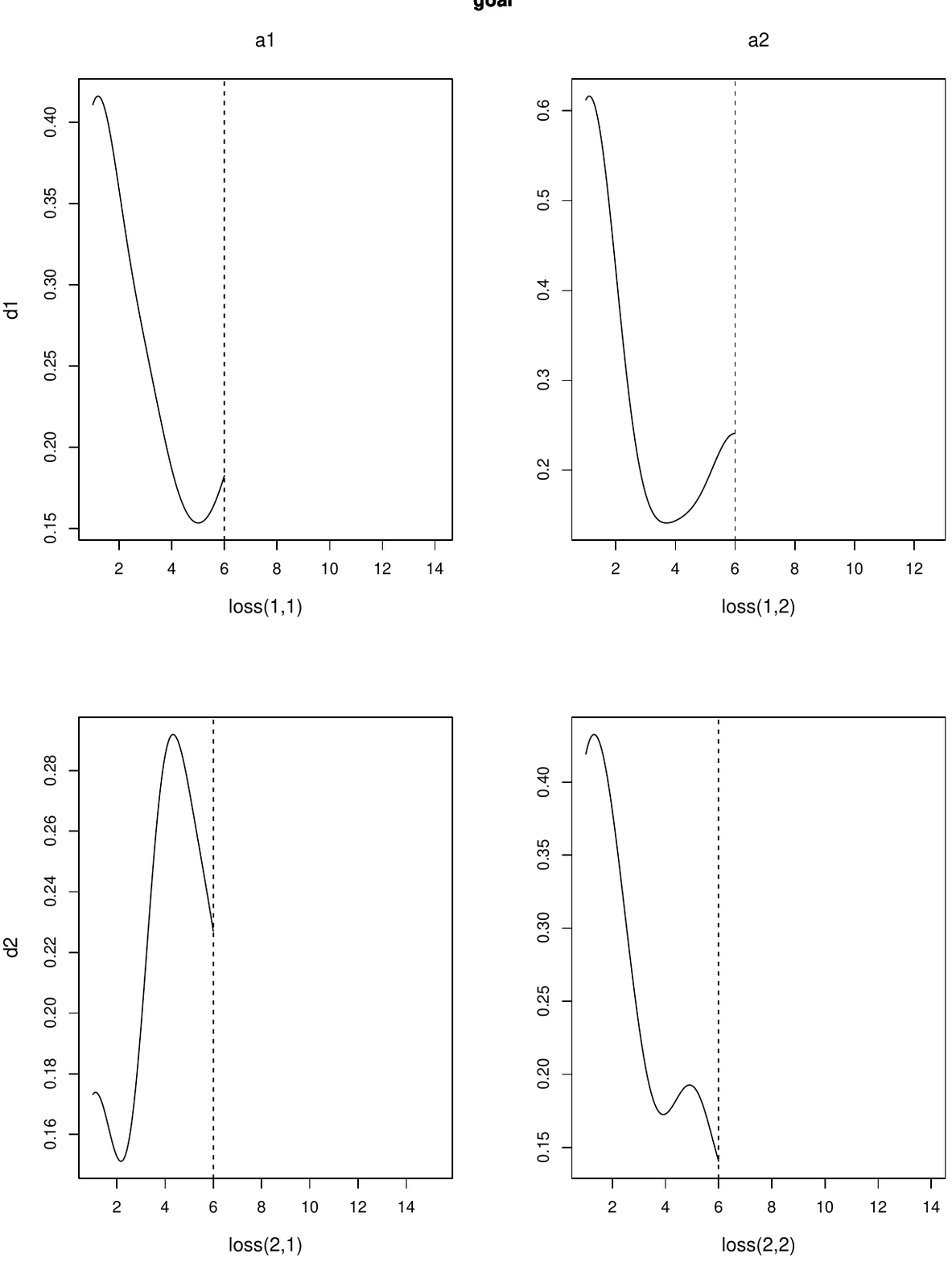}\label{fig:cutoff-6}}
  \caption{Payoff structures with different cutoff points}\label{fig:different-cutoffs}
\end{figure}

So far, we mentioned the necessity of truncating distributions~
only as a technical matter of assuring convergence. As such, the point $a>1$
at which the payoff distributions are truncated (for a compact support) also
influences the outcome of the game, since the equilibria depend on it.
Indeed, the practical choice of $a$ can be made in light of how $a$
determines the $\preceq$-relation among the payoffs.
Informally, Lemma 4.4 in \cite{Rass2015b}~
tells that $\preceq$ is decided based on how the payoff densities behave in a
right neighborhood of $a$. That is, the setting of the value $a$ controls the
range in which damages are considered as relevant for $\preceq$, whereas
damages far lower than $a$ become less and less relevant for the
$\preceq$-relation. This means that the choice of $a$ can be made to
implement a risk prioritization in the model in a sense that is perhaps best
illustrated by an example (cf. also \cite{Rass2015c}~
for a related yet different illustration): 
\begin{exa}\label{exa:risk-prioritization}
Suppose an enterprise has backup capacities to bear losses less than
$50.000~\euro$. Then, we may set the truncation point $a:=50.000$, so that
the entire probability mass assigned to damages $>a$ is ``squeezed''
underneath the distribution on the interval $[1,a]$. Consequently, if a loss
model $F_1$ admits highly likely losses $>50.000~\euro$ (i.e., has in that
sense a fat tail) will result in a higher value of $f_1(a)$ than maybe the
alternative loss distribution $F_2$, assigning smaller likelihood to such
incidents (and in turn coming out with a smaller density $f_2(a)$). Thus,
$f_1(a)>f_2(a)$ will make $F_2$ $\preceq$-preferable over $F_1$
\cite{Rass2015b}.
\end{exa}
The dependence of the resulting equilibrium is best visualized by taking
another example of a $2\times 2$-game with continuous payoff distributions as
shown in Figure \ref{fig:different-cutoffs}. Taking the cutoff point $a=10$
results in a pure equilibrium $(\vec x^*,\vec y^*)=((1,0),(1,0))$, while
truncating the distributions at $a=6$ changes the equilibrium into a mixed
one, namely, $(\vec x^*,\vec y^*)=((0.751,0.249),(0.531,0.469))$. Both of
these have been obtained using fictitious play as described in
\cite{Rass2015b} taking 1000 iterations.

\subsection{Interpretation and Meaning of Equilibria}
The difference in the equilibria computed for the previous examples is
immediately understood by looking at the way how $\preceq$ depends on the
truncated distribution's tails (cf. \cite{Rass2015a,Rass2015b}): if the
cutoff point $a$ is chosen so large that the likelihoods assigned to events
around $a$ are approximately equal, then there is no point in frequently
switching strategies for either player, since the outcome under any of the
four scenarios admitted by the example $2\times 2$-game is roughly the same.
Hence, the equilibrium is ``purified'' by the choice of the cutoff point.
Conversely, if the cutoff point $a$ is chosen so that the masses that
accumulate around it are significantly different, then mixed equilibria may
arise.

This has a twofold and partly positive consequence for the risk manager,
since:
\begin{itemize}
  \item the practical choice of the cutoff point $a$ can be made to mark
      the region in which risks are considered as relevant. That is, if
      there is an agreement that damages beyond some threshold value are
      highly relevant, while everything below it is only of secondary
      interest, then the cutoff point $a$ can be set to exactly this
      threshold value. More importantly, any such setting is not even to be
      taken as sharp, since what determines the $\preceq$-relation is the
      left neighborhood of $a$, so the transition from the domain of
      acceptable risks to the area of critical risks is somewhat smooth.

      For example, if the risk management is done in terms of a CVSS
  scoring, then the risk manager may fix the convention of not caring about
  vulnerability scores in the range around 2, but certainly action is
  demanded for scores above $7$. In that case, we would set the cutoff
  point to $a=7$, in which case the game optimization focuses on scorings
  of 7 or (slightly) less, while the region around scores of $2$ becomes
  relevant only upon equiprobable ratings of all defense actions on the
  range $(2,7]$.

  This effect holds in general and is not limited to the example games
  here. In fact, the cutoff point creates a qualitative difference to
  conventional optimization, where the numeric magnitude of the goal
  function is irrelevant for the maximization. This may not practically be
  the case, but can be accounted for by setting the cutoff point
  accordingly.
  \item An equilibrium must in any case be interpreted \emph{relative} to
      the cutoff point. If $a$ can reasonably/meaningfully be chosen so
      that pure (or approximately pure) equilibria arise, then these
      greatly ease matters of playing the equilibrium. That is, if the risk
      prioritization results in almost all actions ending up with
      approximately equal likelihoods in the relevant neighborhood of $a$,
      then practically playing the optimal behavior is easy, since the best
      response is no longer randomized.
\end{itemize}


\paragraph{Risk vs. Goal Prioritization:}
The role of the cutoff point in risk prioritization must be distinguished
from the role of the weights (discussed in 
\cite[Sec.4.5.1]{Rass2015b}) assigned to reflect the priorities of different
goals in an MGSS. In fact, this is a different degree of freedom, and goal
priorities (the weights $\alpha_i$ for $i=1,2,\ldots, d$ goals can be chosen
independently of the cutoff parameter $a$).

\subsection{Optimizing an Infrastructure's
Resilience (Risk Treatment)}\label{sec:optimization-of-the-infrastructure}


If the line between threats that need to be addressed and those that can be
left aside (for the moment) has been drawn, i.e., the risk evaluation phase
is completed, the risk treatment concerns the selection of controls to
mitigate the threats.

Given one or more threats to be addressed, risk treatment is about selecting
controls against them. The particular selection of controls and resource
allocation is a matter of optimization, and this is where game theory comes
into play. Specifically, it helps to determine the optimal style of
implementing controls towards minimizing the risks. The procedure, on a high
level, is defining a game matrix that has a list $PS_2$ of adversary's
actions (typically the threat or a set of threats to be considered), and
another list $PS_1$ of controls that address the threats. Note that the setup
of the game matrix, specifically the definition of $PS_1$ and $PS_2$ should
be made with care: it may well be the case that some controls in $PS_1$ may
not address certain threats in $PS_2$ and vice versa. For example, if we have
the countermeasures
\begin{itemize}
  \item $d_1\in PS_1$: ``installing an intrusion detection system''
  \item $d_2\in PS_1$: ``install more fire extinguishers''
\end{itemize}
And the threat list $PS_2$ includes
\begin{itemize}
  \item $T_1\in PS_2$: fire outbreak
  \item $T_2\in PS_2$: hacking attack
\end{itemize}
Then it is obvious that $d_1$ is pointless against $T_1$ and $d_2$ has no
mitigating effect on $T_2$. In such cases, one may consider a clustering of
threats and controls based on their mutual relevance. This amounts to
selecting specific controls against specific threats, and allows an optimized
control selection ``per threat'' or ``threat group'' (if a control is
effective against more than one threat). Once the lists $PS_1$ and $PS_2$
have been specified, it pays to sub-classify the actions in $PS_1$, as there
are:

\begin{description}
  \item[Static controls:] these have a permanent effect in being a change
      to the system structure. Their selection must be based on the goal of
      finding the minimal set of controls that covers the maximal spectrum
      of threats. If several candidate controls are available, then the
      problem is ``game-theoretic'' in the sense of being a humble
      optimization for the defender having $n$ strategies against a
      specific single strategy of the attacker (that is the threat to be
      mitigated). If all threats shall be mitigated at the same time, then
      the problem of control selection can be set up as a matrix game,
      where $n$ controls are available against $m$ threats. The resulting
      equilibrium consists of three output artefacts:
      \begin{enumerate}
        \item A probability distribution $\vec{x}^*$ over the $n$
            controls
        \item A probability distribution $\vec{y}^*$ over the $m$ threats
        \item An equilibrium payoff distribution $V$
      \end{enumerate}
	Each of these is meaning- and useful, because:
    \begin{itemize}
      \item All controls having a nonzero probability assigned by the
          equilibrium distribution $\vec x^*$ must be played for risk
          minimization. Thus, these controls need to be implemented. Note
          that the magnitude of the probability is herein not relevant,
          but can be taken as an indication of importance of the
          respective countermeasure. For example, if the game’s
          equilibrium prescribes to ``play'' the strategy ``install
          another firewall'' with likelihood 0.3, then it is obviously
          meaningless to install a firewall only to an extent of 30\%.
          Nonetheless, it tells us that this control must be installed at
          all. However, if the firewall installation gets a likelihood of
          0.3, and another control, say ``change passwords'' gets a
          likelihood of 0.6, then this indicates that both of these
          measures must be implemented, but changing the access
          credentials (or enforcing this) should go first, as it is the
          more frequent (thus more important) strategy in the
          equilibrium.
      \item The likelihoods of different threats can be taken as an
          indication of ``how likely'' a threat is in the worst case. This
          indication must, however, be interpreted in light of the
          non-uniqueness of equilibria, which entails that the computed
          distribution $\vec y^*$ is indeed a worst case scenario, but there
          may be (many) others too.
      \item The equilibrium payoff distribution $V$ quantifies the
          residual risk remaining after the controls have been
          implemented. If, say the average (mean value) of $V$ turns out
          to be not satisfying, then the entire selection process may be
          repeated (in Figure \ref{fig:iso-risk}, this is risk decision
          point 2, where the process may be restarted from the
          ``establish context'' phase.
    \end{itemize}

Alternatively, we can ask for the smallest set of countermeasures that are
simultaneously effective against all threats. Given the information of
which defense measure is effective against which threat (in the form of a
set $S_i$ of candidate countermeasures for the $i$-th threat), the problem
is to compute a \emph{hitting set} for the family $\set{S_1,S_2,\ldots}$.
That is a well-known problem and can be solved using linear optimization,
or more direct algorithms found in the literature.

	  \item[Dynamic controls:] these have a volatile effect and need to be
repeated, such as patching and security awareness trainings. The latter is
also an example that justifies the approach of letting the outcome of a
control being uncertain and in fact quantified by a probability
distribution. In case of a security awareness training, this distribution
could reflect the (relative) amount of people that are (afterwards) highly
aware, medium aware or even remain unaware despite the training. In any
case, the outcome is never certain and depends on the person and its
background. The second example of patching equally well justifies the
modeling of outcomes with probability distributions: patching may be
prescribed, but the patch may be unavailable at the current time, or it may
not help against the right vulnerability. So, the outcome of ``patching''
is also uncertain and the model put forth in \cite{Rass2015a} explicitly
accounts for this. The selection and implementation of such dynamic
controls works as for the static controls described before, but with an
important difference in the use of the artefact distribution $\vec{x}^*$:
this is a randomized prescription telling the frequency of how often a
control is repeated. That is, if a control, say ``patching'', gets assigned
a probability of 60\% by $\vec x^*$, then the implementation is done as
follows: on each day, we toss a biased coin, coming up heads with a chance
of 60\%. If it does so, the administrator is told to look for and install a
patch. The other days, no patching is done. This example is of course
oversimplified, but shall illustrate the ``security-by-randomness''
approach that game theory enforces: security against the attacker is gained
by it not knowing the current patch level reliably (as the patching can
happen every day), so anything that the attacker has been learned so far
may soon be invalidated, and hence the system is more secure.
\end{description}

\paragraph{A Purely Combinatorial Selection Technique:}
A different selection method for risk controls is offered by the methods of
model-based diagnosis (MBD) \cite{Reiter1987}. In a nutshell, MBD (as put
forth in \cite{Reiter1987,Greiner1989}) offers a systematic way of testing
the effectiveness of subsets of $PS_1$ against subsets of $PS_2$. The goal is
finding a minimal set of actions in $PS_1$ that address all threats in
$PS_2$. More formally, let there be a relation $\rightsquigarrow\,\subseteq
PS_1\times PS_2$, with the semantics that
\[
d_1\rightsquigarrow d_2,\text{ if and only if defense }d_1\text{ is working against threat }d_2.
\]
Conversely, if some defense action $d_1$ is useless against threat $d_2$,
then $(d_1,d_2)\notin\,\rightsquigarrow$. Using this relation, we can
associate each threat with countermeasures relevant to it, giving a family of
sets as
\[
    \mathcal{P}:=\set{\set{d_i\in PS_2: d_i\rightsquigarrow t}: t\in PS_1},
\]
i.e., an entry in the family $\mathcal{P}$ is a set $C_i :=
\set{d_{i_1},d_{i_2},\ldots,d_{i_k}}$ of all $k$ countermeasures that are
effective against threat $d_i\in PS_2$.

The sought optimal selection of controls, a \emph{diagnosis} in the
terminology of \cite{Reiter1987}, is a \emph{minimal hitting-set} $D$ for
$\mathcal{P}$. Minimality can here be understood by cardinality or in terms
of $\subseteq$-relation, so that no smaller set than $D$ is a hitting set
(this is the usually preferred setup). By definition, a hitting set $D$ has
the property of intersecting every element in $\mathcal{P}$ (formally, $D\cap
C_i\neq\emptyset$ for all $C_i\in\mathcal{P}$). Again, interpreting this in
our context, this means that the selected security defense measures in $D$
effectively address all threats in $PS_2$ (this is the hitting-set property),
and any proper subset of these controls would leave at least one threat
untouched (this is the $\subseteq$-minimality).

Computing hitting sets is a simple matter of setting up a linear integer
constrained optimization problem, or possible by direct computation and
heuristics. We leave the computational details aside here, except for the
final remark that the problem of computing minimal hitting sets is NP-hard in
general (see \cite{Karp1972}). This is why a game-theoretic method may be
more efficient and hence preferable here.

\subsection{Software Support}\label{sec:tool-support}
%
The entire theory outlined up to this point 
has been implemented as a package for the \texttt{R}-system \cite{RDCT2016},
named \texttt{HyRiM} \cite{Rass2016}. Plots and results given to illustrate
matters in the following have all been obtained using this implementation,
and we will give the respective call sequence (on the \texttt{R} shell) here
in a condensed form that may serve as an example-based tutorial for applying
the theory elsewhere.

It must be emphasized that the implementation mentioned here cannot be a
substitute for the entire risk management process as outlined in section
\ref{sec:risk-man-process}. Duties of context definition, infrastructure
modeling, threat identification and listing of defense actions should be
subject of extensive use of established (relevant) software tools and
standards in this context. Vulnerability scanners like Nessus \cite{TNS2011},
reporting tools like Cauldron \cite{2008}, or tool support to work with the
ISO or BSI standards family are strongly recommended in these initial phases
of the risk management process.

\begin{enumerate}
  \item Assume that the expert opinions are available as a \texttt{numeric}
      vector \texttt{obsij} containing all the answers for the scenario
      $(d_i,d_j)\in PS_1\times PS_2$. That is, \texttt{obsij} is a vector
      of integers in the range $\set{1,2,\ldots,M}$, where $M$ is the
      maximum value (for our examples, we took $M=10$).

      Given a distinct such vector \texttt{obsij} for each $i,j$, we can
  compile a loss distribution by the call:
\begin{verbatim}
fij <- lossDistribution(obsij)
\end{verbatim}
This produces a continuous loss model as displayed in Figure
  \ref{fig:different-cutoffs}. A visualization as in this figure is
  obtained by either plotting a single loss distribution
\begin{verbatim}
plot(f11, cutoff=6)
\end{verbatim}
(in which the parameter cutoff is optional and defaults to the maximum
  observation in the data from which the loss distribution has been
  constructed), or by plotting the entire game matrix by calling:
\begin{verbatim}
plot(G, cutoff=6)
\end{verbatim}
  where the parameter cutoff is again optional and, if supplied, is passed
  to the inner plots of the payoff distributions.

  Assume that the previous step has been repeated to construct continuous
  loss distributions \texttt{f11}, \texttt{f12}, \texttt{f21} and
  \texttt{f22}. Discrete loss models (with smoothing) can be obtained by
  telling \texttt{lossDistribution} to come up with a discrete model. This
  is done by adding an according flag, and (optionally) supply a bandwidth
  parameter (which if omitted, defaults to Silverman's rule of thumb,
  applied to the data is if it were continuous):
\begin{verbatim}
fij <- lossDistribution(obsij,discrete=TRUE,supp=c(1,10))
\end{verbatim}
  The parameter \texttt{supp} is supplied as a reminder (also for the user)
  that all loss distributions need to have the same support (otherwise, the
  construction of games upon a set of distributions with different supports
  will fail with an error message reported).
  \item From the (continuous) loss models \texttt{l11}, \texttt{l12},
      \texttt{l21} and \texttt{l22}, we compile the MGSS by an invocation
      of:
\begin{verbatim}
G <- mosg(n = 2, m = 2,
        losses = list(f11, f12, f21, f22),
        goals=1,
        defensesDescr = c("d1", "d2"),
        attacksDescr = c("a1", "a2"),
        goalDescriptions = c("goal"))
\end{verbatim}
    in which the textual descriptions of the strategy spaces $PS_1$ and
  $PS_2$ are optional and insignificant for the subsequent analysis (they
  only show up in plots).
  \item Given the game object, we can call \texttt{mgss} to give us a
      security strategy. This function takes the cutoff point as a direct
      parameter, which if omitted, defaults to the right end of the
      (common) support of all distributions (no truncation):
\begin{verbatim}
mgss(G,T=1000,cutOff=6)
\end{verbatim}
      Here, the parameter \texttt{T} is the number of iterations to the FP
      algorithm.
\end{enumerate}
The final result returned by \texttt{mgss} is an object of class
\texttt{mosg.equilibrium}, and contains the following fields:
\begin{itemize}
  \item \texttt{\$optimalDefense}: an $n$-dimensional column vector with
      rows labeled by the names of the entries in $PS_1$ (parameter
      \texttt{defensesDescr} in the previous call to \texttt{mosg}).
  \item \texttt{\$optimalAttacks}: an $m$-dimensional column vector with
      rows labeled by the names of the entries in $PS_2$ (parameter
      \texttt{attacksDescr} in the previous call to \texttt{mosg}).
  \item \texttt{\$assurances}: a list of \texttt{mosg.lossDistribution}
      objects (similar as returned by the \texttt{lossDistribution}
      function used to construct the game structure). They can be plotted
      (see Figure \ref{fig:different-cutoffs} for examples), printed in
      detail (using the \texttt{summary} function), and related statistical
      quantities can be worked out such as:
      \begin{itemize}
        \item probabilities for individual attacks (probability mass
            function can be evaluated by the generic function
            \texttt{density})
        \item the cumulative distribution function \texttt{cdf}
        \item moments (of given order, returned by the function
            \texttt{moment}), \texttt{mean}, \texttt{variance} and
            quantiles (at given levels, returned by the generic function
            \texttt{quantile}).
      \end{itemize}
\end{itemize}
For example, after having computed an equilibrium in our generalized setting,
a simple numeric risk measure according to the rule ``risk = damage $\times$
likelihood'' can be recovered by invoking
\begin{verbatim}
mean(eq$assurances$goal)
\end{verbatim}
which gives the assured expected loss ($=$ usual quantitative risk) in the
first security goal (\texttt{g1}) of the game. If, besides the mean, the risk
manager is interested in the expected variation around the mean, then it is
an easy matter of computing the variance of the assured loss distribution as
\begin{verbatim}
variance(eq$assurances$goal)
\end{verbatim}
Finally, if the risk manager is interested in the chances of suffering losses
of categories ``high'' or above, he/she may invoke the \texttt{cdf} function
in the proper way. For example, if category ``high'' has the number 4, then
the chances of getting such a loss or more are given by
\begin{verbatim}
1 - cdf(eq$assurances$goal, 3)
\end{verbatim}
Conversely, computing quantiles tells the losses expected up to a, say 95\%,
chance. That is, calling
\begin{verbatim}
quantile(eq$assurances$goal, 0.95)
\end{verbatim}
tells us the maximal loss (category) that occurs in 95\% of the cases.

\subsection{Playing the Equilibrium Strategies}\label{sec:equilibrium-play}
To cast frequencies from the game into a practical project plan  that tells
the security officer when to take which action, let us fix a particular
action and call $p$ its equilibrium frequency, given the time unit (e.g., one
month). The number of actions per time interval $T$ is therefore $p\cdot T$
and in fact is a Poisson distributed variable with rate parameter $p$. In
turn, the time between two actions is exponentially distributed with
parameter $1/p$. So, if we are at time $t_0$, the next repetition of the
action should be at time $t_0+\Delta$, where $\Delta\sim Exp(1/p)$ is an
exponentially distributed random variable and measured in the unit of time
that was fixed before. Sampling exponentially distributed variables is easy
using uniformly distributed random numbers in the unit interval. Given such a
random value $U\sim Uniform(0,1)$, we simply compute $\Delta=-p\cdot\log⁡(U)$
to have $\Delta\sim Exp(1/p)$.

Note that this method yields the desired fraction of actions (events) per
time unit, which is only correct if the gameplay is presumed with a
periodicity taking time $T$. Thus, the rating of strategies (the
questionnaire; cf. section \ref{sec:expert-surveys}) should be done bearing
in mind the game period $T$.

For example, if the defense is a security awareness training, and the unit of
time $T$ is one month, then the threat assessment could ask something like:
\begin{quote}
Consider an employee whose last security awareness training happened no
more than a month ago. How likely do you think is this person plugging a
virulent USB stick into her/his computer at work?
\end{quote}

A game-theoretic treatment of risks using actions that need repetition will
return likelihoods for each action to occur (in an optimal defense strategy).
Repeatable actions correspond to repeatable and independent game repetitions.
To make this semantically sound, we thus need to make the assessment in light
of the game's periodicity, to decide whether or not it is necessary to take a
particular action again. Returning to the previous example of security
training, we certainly will seek to avoid having such training every month,
but the optimal periodicity of the action is far from obvious. A
game-theoretic analysis will deliver a frequency of training, counted over
the repetitions of the game. If we -- hypothetically -- let the game duration
be one month before the next (hypothetical) iteration starts, then the
question of the effectiveness of a security training must refer to the
current period, which lasts no more than one month up back in time. Hence, we
get the proposed question sketched above.

\section{On Model Validation and Verification}

It lies in the nature of risk management that there is no such thing as a
``rollback'' that could reset a company into the state before an incident, to
verify if the here proposed decision framework or another would have
delivered the better outcomes. The same problem is very well known from other
fields of science, and clinical studies tackle the issue by introducing
control groups to be able to clearly relate observed effects to the treatment
rather than coincidence (or a placebo effect).

Things are no different in the area of risk management, except perhaps at
increased complexity and difficulty. In a famous quote, George Box is telling
us that essentially all models are wrong, but some are still useful. His
point is definitely right, but it must not be taken as an excuse to not ask
for the quality of a decision model. While it is certainly true that
incorrect model parameters lead to incorrect model output, judging the
parameterizations as wrong because of the model performing badly would be
logically flawed (as it incorrectly reverses the implication).

Instead, the model validation must be up to long term studies with control
groups, followed by well-designed statistical tests to reject the hypothesis
of the model performing identically to a chosen alternative. The goal of
rejecting the hypothesis is crucial here, since a statistical test can never
be a formal proof, but only provide an empirical counterexample against some
claim. Thus, to verify the claim of the here-proposed method to be a useful
risk decision supporting framework, we need to refute the claim of it
performing equal to a competing technique. This setting is very well known
from clinical studies incurs some challenges for risk management, since
security incidents are random events, and cannot be considered as independent
or isolated from one another (advanced persistent threats are only one hidden
cause that can manifest itself in a long sequence of seemingly independent
but nevertheless cumulating events up to a final catastrophe). Moreover,
there are no such things as ``lab conditions'', under which the effect of a
decision can be tested against a different one. Thus, a control group in our
setting can only be another company, but incidents -- particularly APTs --
are focused and tailored to the company, so comparing data from one company
to that of another is most likely not too meaningful, since incidents happen
to both companies, but in crucially different forms.

This substantiates the aforementioned long-run of studying the effects of one
decision making against another. Statistical tests are a natural tool here,
but the selection of an appropriate one requires care for several reasons:
\begin{enumerate}
  \item Parametric tests come with assumptions on the distributions of the
      sample sets under investigation. For example, assuming Gaussian
      distributions here could be flawed, since we talk about losses that
      are typically modeled by heavy tailed distributions.
  \item The tested hypothesis must be considered relative to how the risk
      management works. For example, tests on equality of means appear
      quite well-aligned to risk management that minimizes expected
      damages, i.e. means. That is, if the alternative risk management
      against which the proposed method is to be tested considers only
      means (for example, if risk is defined as the product of likelihood
      and impact), while our framework allows for a disregard of this
      quantity (since the hyperreal ordering remains unchanged upon
      disregarding any finite number of moments, which can include the
      first moment), the test may unintendedly be biased towards favouring
      the alternative hypothesis.
\end{enumerate}
Similar issues apply to the selection of the ``control group'', because the
benchmark figures that an enterprise uses to measure its own performance may
be highly different between companies and may be highly specific for the
business area. One benchmark figure that may have wide applicability and --
in the end of the day -- may be the only high-level figure that counts for
risk management, is the \emph{number of severe incidents}. This number
depends on the understanding of ``severe'', and the threshold at which losses
are counted as severe can be defined individually for each enterprise.

Towards a method for validating our risk management decision framework, we
will thus refrain from specifying a particular risk threshold here, and go
for verifying the purely qualitative intuition that:
\begin{quote}
\emph{A good risk management framework should minimize the number of severe
incidents.}
\end{quote}
Consequently, to assess the ``goodness'' of the decision framework based on
game theory, the proposed procedure is as follows, for any \emph{particular}
(i.e., fixed) enterprise:
\begin{enumerate}
  \item Define a control group: since data coming from a different company
      may be biased and relate to different endo- and exogenous processes,
      the control group should be the enterprise itself. This also fixes
      the risk management that we test our method against, since it simply
      is the current decision making framework that we want to judge
      against the new (here proposed) framework.
  \item Fix a meaning of ``severe incident'' in the sense of specifying a
      threshold above which damages or losses count as severe. Given a
      series of historic records on losses that occurred in the past, one
      can compile a past loss distribution $F_{past}$, and, for example,
      define the threshold using a quantile of $F_{past}$. In any case, it
      is important to fix this threshold a-priori (still, we need to bear
      in mind that the thresholds may need adaption over the lifetime of
      the enterprise to remain meaningful), and independently of the value
      implied by \cite[Thm.2.14]{Rass2015a}.
      (this is to avoid a bias in favour of the here proposed decision
      making technique). Call the threshold $t_0$, and let us think of
      losses $>t_0$ as being \emph{severe}.
  \item Collect the number $n_0$ of incidents whose losses were $\geq t_0$,
      i.e., which have been severe in this understanding. Based on the
      distribution $F_{past}$ of past losses, the likelihood for such an
      incident to occur would be $p=\Pr(L\geq t_0)$, if $L$ measures the
      random loss. Assuming independence of events, the \emph{number $N$ of
      severe past incidents} is Poisson-distributed with a rate parameter
      $\lambda_{past}$ that depends on $p$.

      For the data collection as such, the independence assumption entails
      that common underlying causes of seemingly independent events must be
      identified before feeding this data into a test statistic. Several
      events with a single common cause must then be compiled into a single
      event (representing the underlying cause) with a total loss incurred.
      For example, if many events can be related to an APT, the typical
      nature of these symptoms is to stay undetected, so they would
      probably not count into the set of ``severe'' incidents. Nonetheless,
      if the APT manifests itself in several quite severe damage events,
      then these must be considered as \emph{one single} incident (being
      the APT, only exhibiting multiple damages) with accumulated effects.
  \item At some point in time, the new risk management framework (not
      necessarily the one described in this report) becomes installed, and
      a new recording of ``severe'' events (with the same definition as was
      used in the past) must start. Call the number of these $M$, and
      observe that the new loss distribution $F_{new}$ is as well Possonian
      with a rate parameter $\lambda_{new}$, which depends on the mass $p$
      that $F_{new}$ assigns to the region $\geq t_0$.
  \item Fix equal time periods for data collection: For the new collection
      of losses, it is crucial that the time period over which $M$ was
      counted \emph{equals} the period over which $N$ was taken. Otherwise,
      the test would be biased, since if the past period is much shorter
      than the current one, then the number of past incidents is
      necessarily much smaller than the number of current incidents.
      Likewise, if the past period was very long, and the new period is
      relatively short, then the number of incidents $N$ will necessarily
      be much higher than the number $M$ of currently observed incidents.
  \item A final assumption that must nevertheless be emphasized is both
      rates $\lambda_{past}, \lambda_{new}$ being taken as \emph{constant}
      over time (otherwise, the events would not be Poisson distributed any
      more).
\end{enumerate}

Given the past and current samples with sizes $n$ and $m$, respectively, the
distributions of $N$ and $M$ are $N\sim\text{Poisson}(n\lambda_{past})$ and
$M\sim \text{Poisson}(m\lambda_{new})$. The hypothesis to be refuted is the
past risk management performing better or equal to the new method, in which
case the null hypothesis becomes
\[
    H_0: \frac{\lambda_{past}}{\lambda_{new}}\leq 1,
\]
which expresses that the past risk management reduces severe incidents at
least as good as the current (new) one, hence the new risk management is no
better than the previous.

The statistical test of choice is a one-sided rate ratio test
\cite{Gu&Ng2008,NCSSSoftware2016}. These references discuss various methods
to test Poisson rates, and one exact test is directly available in \texttt{R}
via the package \texttt{rateratio.test} \cite{Fay2010,Fay2015}, which can be
invoked by issuing the command:
\begin{quote}
\texttt{rateratio.test(c(}$n\cdot \lambda_{past}$\texttt{,} $m\cdot
\lambda_{new}$\texttt{), c(}$m$\texttt{,} $n$\texttt{), alternative =
"greater")}
\end{quote}
for our case. The null-hypothesis is rejected if the $p$-value is less than
the pre-specified $1-\alpha$ (that defaults to
$\alpha=$\texttt{conf.level}$=0.95$ in the above call).

A general validation or verification of the method against competing ones is
thus a matter of several companies running long-time evaluations and deciding
on their own if the decision framework proposed here performs better or worse
than their competing candidate. Extending the investigation towards checking
whether or not the stochastic order makes up a good model of decision making
under bounded rationality is an independent challenge and widely extended
line of research that we will not go farther into here. Instead, the last
test is meant to complete the theory developed here, and we close the report
by a discussion in favour of the qualitative advantages that this form of
decision making may offer for practitioners.

%

\section{Conclusion and Outlook}

The most important pros of the method can be roughly summarized as follows:
\begin{itemize}
  \item \emph{No consensus problem}: The approach of working with entire
      distributions rather than single values avoids issues of consensus
      finding among experts. That is, if there are disagreeing opinions
      among a group of people, every opinion goes into the risk analysis at
      equal weight. Thus, nobody has to be overruled or otherwise accept a
      compromise that deviates from the own thoughts. If, subsequently,
      damage occurs nonetheless, arguments of people claiming to have
      foreseen this incident are useless, since any such warning was
      considered from the beginning on already.
  \item \emph{Vague assessments allowed}: Even for a single expert, there
      is no need to confine oneself to a single risk assessment, perhaps at
      the risk of being mistakenly understood as also saying what the risk
      ``is not''. In other words, if a security expert utters a risk of
      level ``medium'', then this is not the implicit additional statement
      that the risk is ``not high''. In fact, in allowing the expert to add
      uncertainty to her/his assessment\footnote{either explicitly by
      giving the loss distribution directly (cf. Example
      \ref{exa:simple-modeling}), or implicitly by adjusting the bandwidth
      parameter of the kernel smoothing (see Section
      \ref{sec:discrete-kernel-smoothing} and Example
      \ref{exa:simple-modeling}) to assign likelihoods to other
      possibilities around the actual assessment.}, the intrinsic fuzziness
      of perceived security risks can be (conveniently) expressed.

      This degree of freedom also helps to treat scenarios in which several
  threats occur jointly. If a threat scenario is such that most likely
  other attacks are mounted along, then the expert may still focus on the
  ``main objective'' (main threat) of the assessment, while explicitly
  stating and rating the possibility of more severe damages upon the
  simultaneous occurrence of other threats by assigning some likelihood to
  the other alternatives (cf. remark
  \ref{rem:multiple-simultaneous-threats}).
  \item \emph{Enforced systematization}: The need to set up a game matrix
      for all combinations of defenses and attacks enforces a systematic
      approach to the entire risk management problem. The method itself is
      herein merely embedded and rests on a comprehensive threat analysis
      and identification of countermeasures. The method of risk assessment
      as outlined in section \ref{sec:expert-surveys} is actually an
      application of \emph{crowdsourcing}; see \cite{Bratvold2016}.
\end{itemize}
Perhaps the most important advantage over more conventional quantitative
methods of risk assessment is the following: none of the information items
(except the average) returned by the call sequences at the end of Section
\ref{sec:tool-support}, can be obtained by standard techniques of risk
management that are entirely based on expected losses as risk. Hence, the
preservation of information throughout the entire process from the modeling
up to the final risk result interpretation is an important benefit over other
methods. In brief: the here proposed method is preserving information of all
experts over the entire risk management process, and avoids loss of
information through consensus finding and opinion or risk aggregation.

\section*{Acknowledgment}
This work was supported by the European Commission's Project No. 608090,
HyRiM (Hybrid Risk Management for Utility Networks) under the 7th Framework
Programme (FP7-SEC-2013-1). The author is indebted to the whole consortium of
HyRiM, and particular gratitude is expressed to Sandra K\"{o}nig and Stefan
Schauer from the Austrian Institute of Technology for invaluable discussions
and suggestions along this entire research.

\bibliographystyle{plain}

\end{document}